\documentclass[preprint2]{aastex} 
\usepackage{graphicx} 
\slugcomment{Submitted to the Astronomical Journal } 

\shorttitle{Shocked Cloud in the Cygnus Loop SNR} 
\shortauthors{Fesen et al.} 

\begin{document} 
\acknowledgments 

\title{An Isolated, Recently Shocked ISM Cloud in the Cygnus Loop
SNR } 

\author{D.J. Patnaude\altaffilmark{1}, R.A.
Fesen\altaffilmark{1}, 
J.C. Raymond\altaffilmark{2}, N.A. Levenson\altaffilmark{3}, 
J.R. Graham\altaffilmark{4}, 
and Debra J. Wallace\altaffilmark{5}} 
\altaffiltext{1}{6127 Wilder Laboratory, Physics \& Astronomy
Department, 
Dartmouth College, Hanover, NH 03755} 
\altaffiltext{2}{Harvard-Smithsonian Center for Astrophysics, 
60 Garden Street, Cambridge, MA 02138} 
\altaffiltext{3}{Department of Physics and Astronomy, 
University of Kentucky, 
Lexington, KY 40507} 
\altaffiltext{4}{Department of Astronomy, 
University of California, Berkeley, 
Berkeley, CA 94720} 
\altaffiltext{5}{Department of 
Physics and Astronomy, Georgia State University, 1 University Plaza, 
Atlanta, Georgia, 30303} 

\begin{abstract} 

Spatially resolved {\it ROSAT} X-ray and ground-based optical data for 
the southwestern region of the Cygnus Loop SNR reveal in unprecedented detail 
the very early stages of a blast wave interaction with an isolated 
interstellar 
cloud. Numerous internal cloud shock fronts near 
the upstream flow and along the cloud edges are visible optically 
as sharp filaments of enhanced H$\alpha$ emission. Faint X-ray emission 
is seen along a line of Balmer-dominated shock filaments 
north and south of the cloud with an estimated X-ray gas temperature of
$1.2 \times 10^{6}$ K (0.11 keV) corresponding to a shock velocity of 
290 km s$^{-1}$.
The main cloud body itself exhibits 
little or no X-ray flux. Instead, X-ray emission is confined along the 
northern and southernmost cloud edges, with the emission brightest in the 
downstream regions farthest from the shock front's current position. 
We estimate an interaction age of $\sim$ 1200 yr based on the 
observed shock/cloud morphology.

Overall, the optical and X-ray properties of this shocked ISM cloud show 
many of the principal features predicted for a young SNR shock -- ISM 
cloud interaction. In particular, one sees shocklet formation and 
diffraction inside the inhomogenous cloud along with partial main blast 
wave engulfment. However, several significant differences 
from model predictions are also present including no evidence for 
turbulence along cloud edges, diffuse rather than filamentary 
[\ion{O}{3}] emission within the main body of the cloud, unusually 
strong downstream [\ion{S}{2}] emission in the postshock cloud regions, 
and confinement of X-ray emission to the cloud's outer boundaries. 

\end{abstract} 
\keywords{ISM: individual (Cygnus Loop) - supernova remnants -
ISM: emission 
lines and dynamics } 

\section{INTRODUCTION} 

Supernova remnants (SNRs) shape and enrich the chemical and dynamical 
structure of the interstellar medium (ISM) which, in turn, affect the 
evolution of a SNR. Knowledge of just how SN generated shock waves 
travel through and interact with the ISM and interstellar clouds is 
fundamental to our understanding of the emission and dynamical details 
of this process. 

Because of its large angular size (2.8$^{\circ} \times$ 3.5$^{\circ}$), 
low foreground extinction ($E[B-V] = 0.08$ mag; 
\citealt{parker67,fesen82}), and wide range of shock conditions, the 
Cygnus Loop is one of the best laboratories for studying the ISM shock 
physics of middle-aged remnants. At a distance of 440$_{-110}^{+150}$ pc 
\citep{blair99}, it has a physical size of 21 $\times$ 27 pc. Located 
8.5$^{\circ}$ below the galactic plane, the Cygnus Loop lies in a 
multi-phase medium containing large ISM clouds with a hydrogen density 
of $n=5-10$ cm$^{-3}$, surrounded by a lower density intercloud 
component of $n\approx0.1-0.2$ cm$^{-3}$ \citep{denoyer75}. 

The currently accepted view of the Cygnus Loop is that it represents an 
ISM cavity explosion by a fairly massive progenitor star 
\citep{mccray79,charles85,levenson99}. The cavity is presumably the 
result of strong stellar winds emanating from the high-mass progenitor. 
In this picture, the supernova shock has been traveling relatively 
unimpeded for a distance $\approx$ 10 pc and has only relatively 
recently begun to reach the cavity walls. The interaction of the shock 
with the cavity walls is responsible for the remnant's observed radio, 
optical, and X-ray emission. 

Previous studies of the Cygnus Loop have examined selected regions in 
the UV/optical and X-ray 
\citep{ku84,hester86b,graham95,levenson96,danforth00}. These have shown 
that there are two distinctly different types of optical line-emission 
filaments present. The Cygnus Loop's brighter filaments are the result 
of shocked and subsequently radiatively cooled interstellar clouds whose 
preshock densities are many times that of the intercloud regions. Along 
with hydrogen and helium recombination line emissions, these filaments 
exhibit strong forbidden line emissions from oxygen, nitrogen, and 
sulfur, and are located downstream from the advancing shock front in 
postshock gas with temperatures $\sim$ 10$^{5}$ K \citep{fesen85}. The 
degree of postshock cooling (``incompleteness'') can strongly affect 
the relative strength of the line emissions, particularly the observed 
[\ion{O}{3}] $\lambda\lambda$5007,4959 vs. H$\beta$ emissions. In the 
case of the Cygnus Loop, like most other evolved SNRs, comparisons with 
model calculations show its bright filaments have shock velocities 
$\approx$100 km s$^{-1}$. 

Fainter, so-called Balmer-dominated filaments result when a 
high-velocity shock encounters partially neutral gas 
\citep{chevalier78,chevalier80}. The collisionless shock accelerates and 
heats interstellar ions and electrons through electromagnetic plasma 
instabilities, while leaving neutral atoms unaffected. Subsequently, 
the neutral atoms, particularly neutral hydrogen, can be collisionally 
excited as well as collisionally ionized thus permitting the emission of 
Balmer photons with a narrow line profile width corresponding to the 
preshock gas temperature of T $\sim 5000 - 10000$ K. However, neutral 
hydrogen in the postshock region can also undergo charge transfer 
thereby acquiring thermal energy and flow velocities similar to those of 
the shocked ions. Consequently, charge transfer to the shock-heated 
protons produces fast moving hydrogen atoms, which will emit Balmer 
photons with broad line profiles upon collisional excitation. 

Other elements are also collisionally ionized and may also emit line 
photons. However, in the case of neutral atoms and relatively 
low-ionization ions, a line's luminosity is proportional to its 
ionization time, collisional excitation rate, and elemental abundance. 
This leads to relatively weak metallic lines compared to the hydrogen 
Balmer lines and thus Balmer-dominated filaments. 

In general, lower density intercloud regions of the remnant experience
higher velocity shocks and correspondingly show higher postshock 
temperatures. These intercloud regions are thus responsible for a 
remnant's X-ray and coronal line emissions \citep{mckee75,ku84,teske90}. 
X-ray analyzes provide information on elemental abundances, shock front 
position, grain destruction, and other properties of the postshock gas. 
Furthermore, in cases where the postshock gas is fully ionized, X-ray 
emission can yield a direct measure of the shock front velocity. 

Attempts to fit shock models to the observed optical line emission seen 
throughout the Cygnus Loop have sometimes been hindered by surprisingly 
large [\ion{O}{3}]/H$\beta$ ratios leading to the notion of incomplete 
shock emission \citep{blair91}. Using IUE observations, 
\citet{raymond80} found that much of the hydrogen recombination zone 
predicted by steady-flow models is absent, implying that the interaction 
is fairly young, with an incomplete postshock cooling and recombination 
zone. Similarly, small portions of NE limb Balmer-dominated filaments 
have been found to exhibit incomplete postshock cooling zones, 
apparently marking locations of increased ISM density and hence 
somewhat shorter cooling times. 

Some of the results from these analyses are likely affected by other 
factors including limb projection effects, uncertain location of the 
associated forward shock, and the superposition of multiple shock fronts 
along the line of sight. Ideally, to compare observations with model 
emission calculations, one would like to observe a a single, isolated 
ISM cloud largely free of such complicating effects. 

Towards this goal, \citet{graham95} combined X-ray and optical data to 
study a small cloud in the southeast \citep{fesen92} seen in the early 
stages of shock interaction. They found that the Balmer dominated 
emission, together with X-ray emission, traced out the shock front as it 
wrapped around the cloud. Their analysis together with optical 
images taken with the Hubble Space Telescope \citep{levenson01} led 
them to picture the cloud, initially identified as small and isolated, 
as in fact an extension of a much larger cloud, which the blast wave is 
just now interacting with. A complex morphology of interacting shock 
fronts is seen where sharp filaments mark regions where the shock front 
is viewed edge-on and diffuse emission where the view is face-on. 

Here we present optical and X-ray data on a small, isolated and 
recently shocked cloud located along the southwestern limb of the 
remnant. The shock-cloud interaction is viewed nearly edge-on, with the 
shock front visible both within and around the cloud. Using ground-based 
optical and ROSAT X-ray images and spectra, we present an analysis of 
the very early stages of this shock-cloud interaction. Our 
optical and X-ray observations are described in \S\ref{sect:observ}. 
We discuss the results in \S\ref{sect:res} and compare these properties 
to those of other previously studied regions of the Cygnus Loop in 
\S\ref{sect:disc}. In \S\ref{sect:conc}, we summarize our results and 
discuss its implications for the modeling os shock-cloud dynamics. 

\section{OBSERVATIONS} 
\label{sect:observ} 

\subsection{Optical Images and Spectra} 

Narrow-band images of the southwest region of the Cygnus Loop were 
obtained in July 1992 using both the MDM 2.4~m Hiltner and 1.3~m 
McGraw-Hill telescopes. Four 600~s H$\alpha$ filter (FWHM = 80 \AA) 
exposures were taken using the 2.4~m telescope with a Loral 2048 
$\times$ 2048 CCD. These images had a scale of $0 \farcs 343$ 
pixel$^{-1}$. Three 1200~s H$\alpha$ filter images using a narrower 
filter (FWHM = 25 \AA) were also obtained using the 1.3~m telescope with 
the same Loral CCD which yielded an image scale of $0\farcs625$ 
~pixel$^{-1}$. Wider field images taken in H$\alpha$ were subsequently 
obtained in September 1992 using the KPNO 0.6~m Schmidt telescope with 
the S2KA 2048 $\times$ 2048 CCD and a FWHM = 12 \AA \ H$\alpha$ filter. 
This system provided a 68\arcmin~$\times$ 68\arcmin~field of view and a 
resolution of $2 \farcs 03$ pixel$^{-1}$. The S2KA chip suffered from 
two broad, bad pixel columns which were removed using neighboring pixel 
replacement techniques. These patched bad pixel regions can be seen as 
blurred lines in the reduced images. 

Additional line emission images of the Cygnus Loop's southwest region 
were taken in June 1993 using the MDM 1.3~m telescope and H$\alpha$, 
[\ion{O}{3}], [\ion{S}{2}], and [\ion{O}{1}] filters. These filters had 
FWHM bandpasses of 25, 100, 80, and 40 \AA~ respectively. Images taken 
through these filter were obtained using a Tektronix 1024 $\times$ 1024 CCD 
yielding a resolution of $0 \farcs 51$ pixel$^{-1}$. Exposure times were 
$3 \times 600$ s for H$\alpha$, $4 \times 1200$ s for [\ion{O}{3}], 
$3 \times 1200$ s for [\ion{S}{2}], and $3 \times 600$ s for 
[\ion{O}{1}].

East-west aligned, long-slit spectra were then obtained in 1993 
September at two locations in the southwest cloud using the MDM 1.3~m 
telescope. The spectra were taken with the Mark~III spectrograph using a 
5800 \AA/600 lines mm$^{-1}$ grism and a Tektronix 1024 $\times$ 1024 
CCD. This combination yielded an effective slit scale of 
$0 \farcs 77 $ pixel$^{-1}$, with a slit length of approximately 
5\arcmin~ width of $1 \farcs 5$, and spectral resolution of 
$\lambda/\Delta \lambda \approx 1200$. 

The optical CCD images and spectra were reduced using 
IRAF\footnote{IRAF is distributed by the National Optical Astronomy 
Observatories (NOAO), which is operated by the Association of 
Universities for Research in Astronomy, Inc. (AURA) under cooperative 
agreement with the National Science Foundation.} software. The data were 
bias subtracted, flat-fielded, and cosmic-ray hits were removed. The 
two-dimensional spectra were also background and sky subtracted and then 
flux calibrated using observations of the standard stars Kopf 27 and 
BD+25-3941. One-dimensional spectra were then extracted for select areas 
of each slit position. 

The two-dimensional spectra were then used to derive approximate flux 
calibrations of the optical images. Average H$\alpha$, [\ion{O}{3}], 
[\ion{S}{2}], and [\ion{O}{1}] fluxes were obtained from the extracted 
one-dimensional spectra and compared to the total background subtracted 
counts in the same regions of the optical image. The ratios were then 
averaged and the optical images were multiplied by the scaling factor. 
Since count rates vary substantially in different regions of the cloud, 
the derived fluxes are probably not accurate to better than 20\%. 

\subsection{X-ray Images} 

The entire southwest portion of the Cygnus Loop was observed with the 
{\it ROSAT} Position Sensitive Proportional Counter (PSPC). The observation 
was performed in April 1994 for 14029~s with a target center of 
$\alpha$(2000) = 20$^{\rm h}$ 47$^{\rm m}$ 52$^{\rm s}$, $\delta$(2000) = 
29\degr ~17\arcmin ~41\arcsec. 

The {\it ROSAT} PSPC is sensitive to photons in the 0.1--2.4 keV energy 
range, and has moderate energy resolution ($E/\Delta E \sim 2$ at $\sim$ 1 
keV) with spatial resolution better than 30\arcsec~(FWHM on-axis). The 
spatial resolution decreases with increasing off-axis angle becoming 
mirror-dominated when the off-axis angle is greater than $\sim10'$ and 
reaches a value of several arcminutes at the edge of the 2\degr~field of 
view. The mirror assembly is subject to vignetting producing a decrease 
in effective area with increasing off-axis angle. This effect is larger 
at higher energies. Apart from the edges, the PSPC energy resolution is 
essentially uniform over the entire field of view (see 
\citealt{trumper83} for further details on the {\it ROSAT} mission). 

Various checks were performed to verify the quality of the data covering 
the southwest portion of the Cygnus Loop, as produced by the Standard 
Analysis Software System (SASS; version 5.3.2). In particular, we 
checked the overall rejection efficiency against spurious particle 
events, as well as the possible contribution of solar scattered X-rays 
not being completely screened out. The removal of this contamination is 
particularly important in the analysis of diffuse emission like that of 
an extended SNR, as discussed by \citet{snowden93}. 

In undertaking this task, we took advantage of the calibration of 
particle PSPC background by \citet{snowden92} and from the complete 
model of the scattered solar X-ray photons reported by 
\citet{snowden94}. Following \citet{snowden92} and \citet{snowden93}, we 
predict that based on the master veto rate of anti-coincidence counters 
and the Sun-Earth-satellite angle, the best estimate of background 
intensity for our observation is $\sim$ 1.7 $\times$ 10$^{-3}$ counts 
s$^{-1}$ arcmin$^{-2}$. This value compares well with the measured value 
of $\sim$ 1.5 $\times$ 10$^{-3}$ counts s$^{-1}$ arcmin$^{-2}$ and 
indicates that our data are not heavily contaminated. 

{\it ROSAT} data are also affected by other known problems. Of particular 
concern is the occurrence of electronic ghost images for counts in the 
first two SASS channels ($<$ 0.11 keV). This is due to the inability of 
low-energy events to properly trigger the processing electronics which 
computes their  locations \citep{nousek93}. Various soft-energy 
contaminants (e.g., bright Earth or internal background) and after pulse 
events \citep{snowden93} are also of concern. To remove any such 
complicating effects, we discarded all events below 0.11 keV. 

In order to model the X-ray emission from the southwest portion of the 
Cygnus Loop in and around the cloud, we chose to fit the extracted 
spectra from various regions, thereby providing us with a temperature 
map with 3\arcmin~$\times$ 5\arcmin~resolution. We fit the data to a 
single-temperature Raymond-Smith thermal model for an optically thin 
plasma with cosmic abundances \citep{raymond77}. We adjusted three free 
parameters in this model, namely, the temperature, the normalization 
(i.e. the plasma emission measure if the source distance is known), and 
the interstellar hydrogen column density, $N_{\rm H}$, as a parameterization 
of X-ray absorption by the ISM as described by \citet{morrison83}.

The data was fit using the HEASARC XSPEC V11.0 package. Given that the 
data are not Gaussian ($<$ 25 counts per bin), the data must be weighted 
appropriately. We chose to use the Churazov weighting scheme 
\citep{churazov96}, whereby the weight for a given data channel is 
estimated by averaging the counts in surrounding channels. For each 
region we fit, several trials were attempted to avoid convergence to a 
local minima. In particular, we first fit N$_{\rm H}$ and kT, and then varied 
them over a finite sized grid to search out the best $\chi^2$. Finally, 
90\% confidence intervals were calculated for each best fit parameter.

\section{RESULTS} 
\label{sect:res} 

\subsection{Optical Images} 

\begin{figure} 
\includegraphics[bb=20 20 575 455,width=6in,height=6in,keepaspectratio=true]{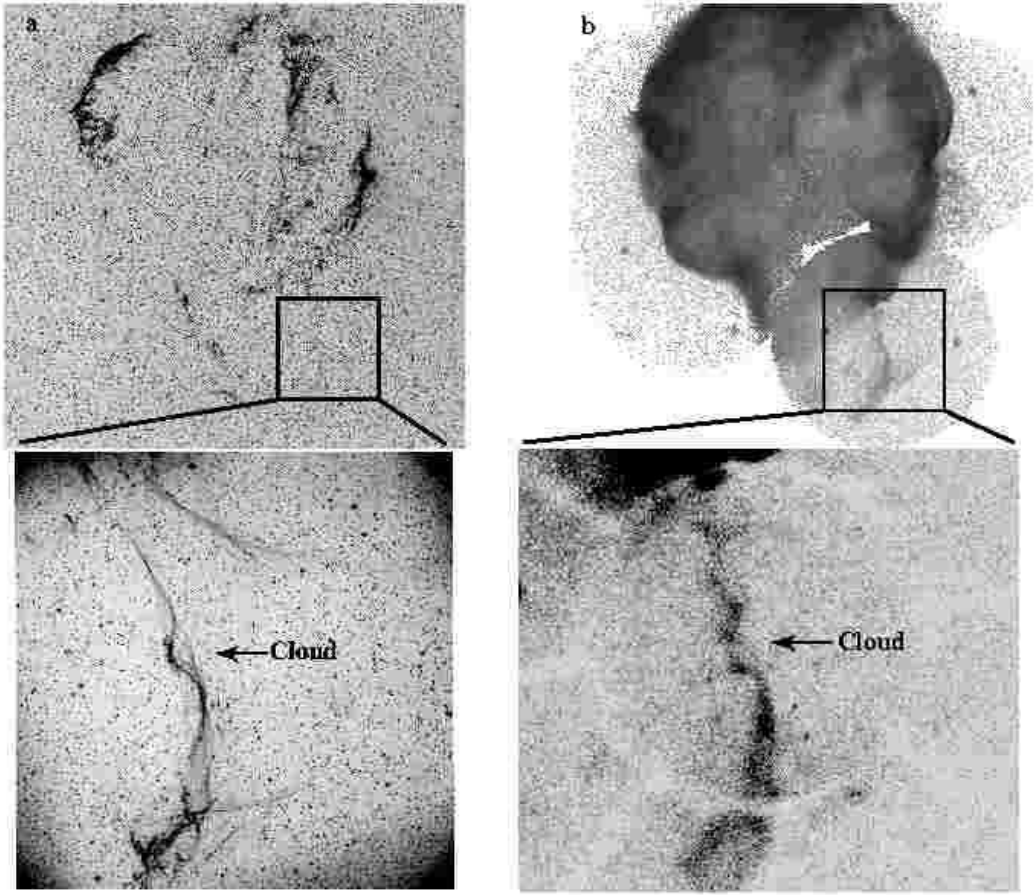}
\caption{{\it a} ({\it top}): Digital Sky Survey H$\alpha$ image of the
Cygnus Loop showing the 
location of the southwest cloud. ({\it bottom}): 
KPNO Schmidt image of the southwest 
cloud. 
{\it b} ({\it top}): {\it ROSAT} PSPC mosaic of the Cygnus Loop in
the 0.2 -- 2.4 
keV band. ({\it bottom}): {\it ROSAT} PSPC image of the southwest
region of 
the Cygnus Loop. Both images have been logarithmically scaled.} 
\label{fig:schmidtpspc} 
\end{figure}

The southwest cloud is located along the western limb of the southern
``breakout'' region of the Cygnus Loop. 
Figure~\ref{fig:schmidtpspc} shows the Digital Sky Survey (DSS)
image of the whole remnant with 
a blow-up of the southwest section imaged in H$\alpha$ using the
KPNO Schmidt 
(left-hand panels). {\it ROSAT} PSPC images of the whole remnant 
and the 14 ksec pointing PSPC 
image are also shown for comparison (right-hand panels). 

\begin{figure} 
\includegraphics[bb=20 20 575 502,width=6in,height=6in,keepaspectratio=true]{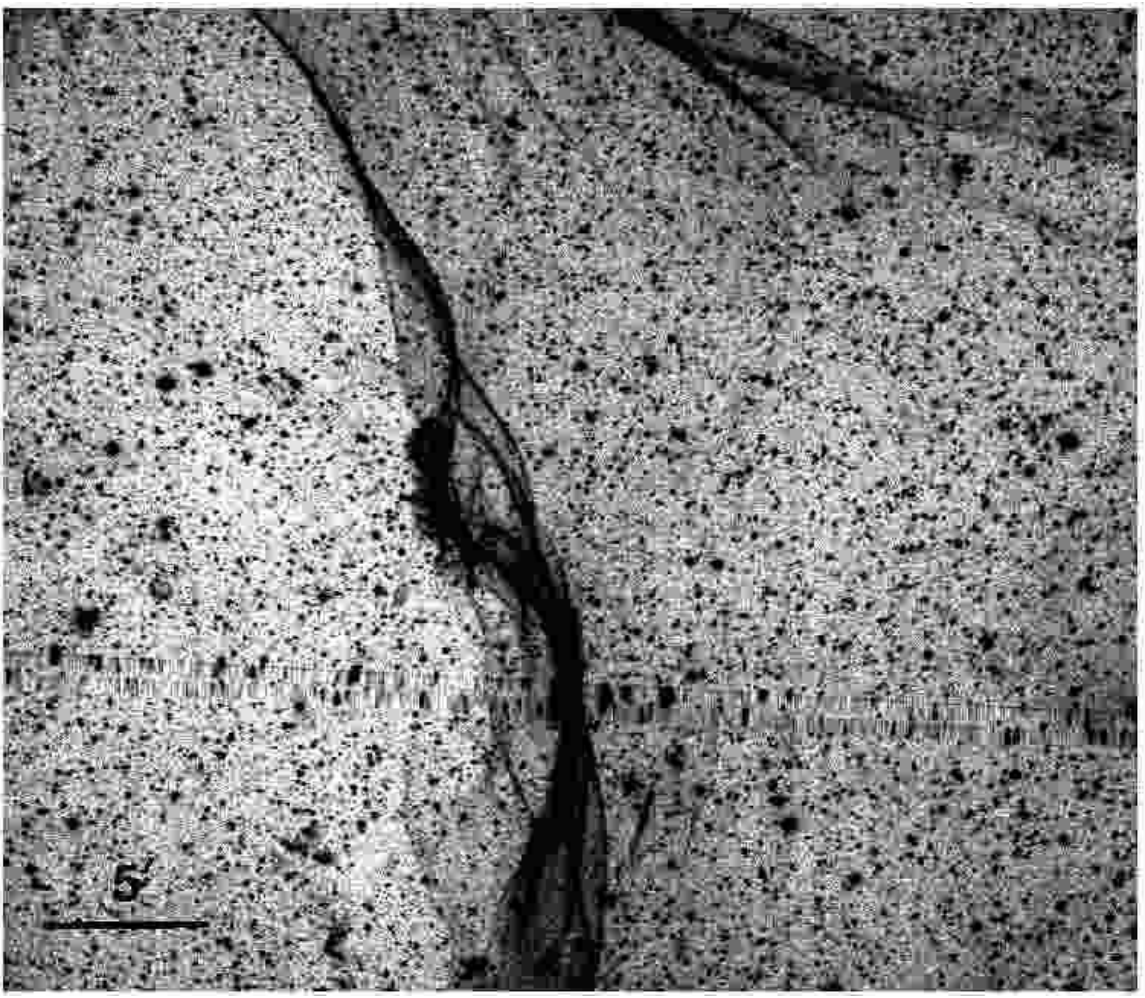}
\caption{KPNO Schmidt H$\alpha$ image of the southwest cloud and
vicinity. 
North is up and east is to the left.} 
\label{fig:schmidt} 
\end{figure}

The morphology of the southwest cloud and vicinity in H$\alpha$ 
is complex in detail yet 
possesses a relatively simple underlying geometry. 
This can be seen in Figure~\ref{fig:schmidt} which shows an
enlargement of 
our KPNO Schmidt H$\alpha$ image of the cloud and vicinity. 
The cloud itself measures 1\arcmin~$\times$~3\arcmin~(0.1
$\times$ 0.3 
pc) and coincides with a marked indentation along a $\sim$ 40$'$
long 
H$\alpha$ filament. 
Sharp filaments appear to mark the dividing line between 
a partially ionized medium -- seen here as faint diffuse emission 
ahead (west) of the shock -- and fully ionized, shocked interstellar
gas behind 
and to the east of the shock fronts. 
This partial/complete ionization demarcation is complex due in part
to the 
the shock front's interaction with the southwest cloud, 
which has created a series of slower moving shock fronts draped
around the 
cloud. 
Because these lag behind the undisturbed shock front, 
some preshock diffuse emission can be seen to lie behind the main
shock 
front; 
for example, in the region just north of the cloud. Here, the 
emission behind the westernmost shock front is 
partially pre-shock emission seen in projection. 
As expected, this faint emission vanishes at the edge of the 
trailing (eastern) shock front. 
A similar situation can be seen $\sim$ 10\arcmin~south of the cloud, 
where a portion of the shock front is more severely distorted 
resulting in somewhat larger projection effects. 

The overall impression one gets from examination of the whole 
region (Figs.~\ref{fig:schmidtpspc} \&~\ref{fig:schmidt}) 
is of a corrugated shock front with large 
distortions due to ISM density variations on both large and small
scales. 
Indeed, the presence of H$\alpha$ filaments running 
almost E-W along both the top and bottom of the Schmidt image 
(Fig.~\ref{fig:schmidtpspc}) suggests the presence 
yet another a shock breakout in this 
general direction. The nature of these nearly E-W running filaments 
is beyond the scope of this paper and we will 
concentrate here on just the southwest cloud and immediate
surroundings. 

\begin{figure} 
\includegraphics[bb=20 20 575 565,width=6in,height=6in,keepaspectratio=true]{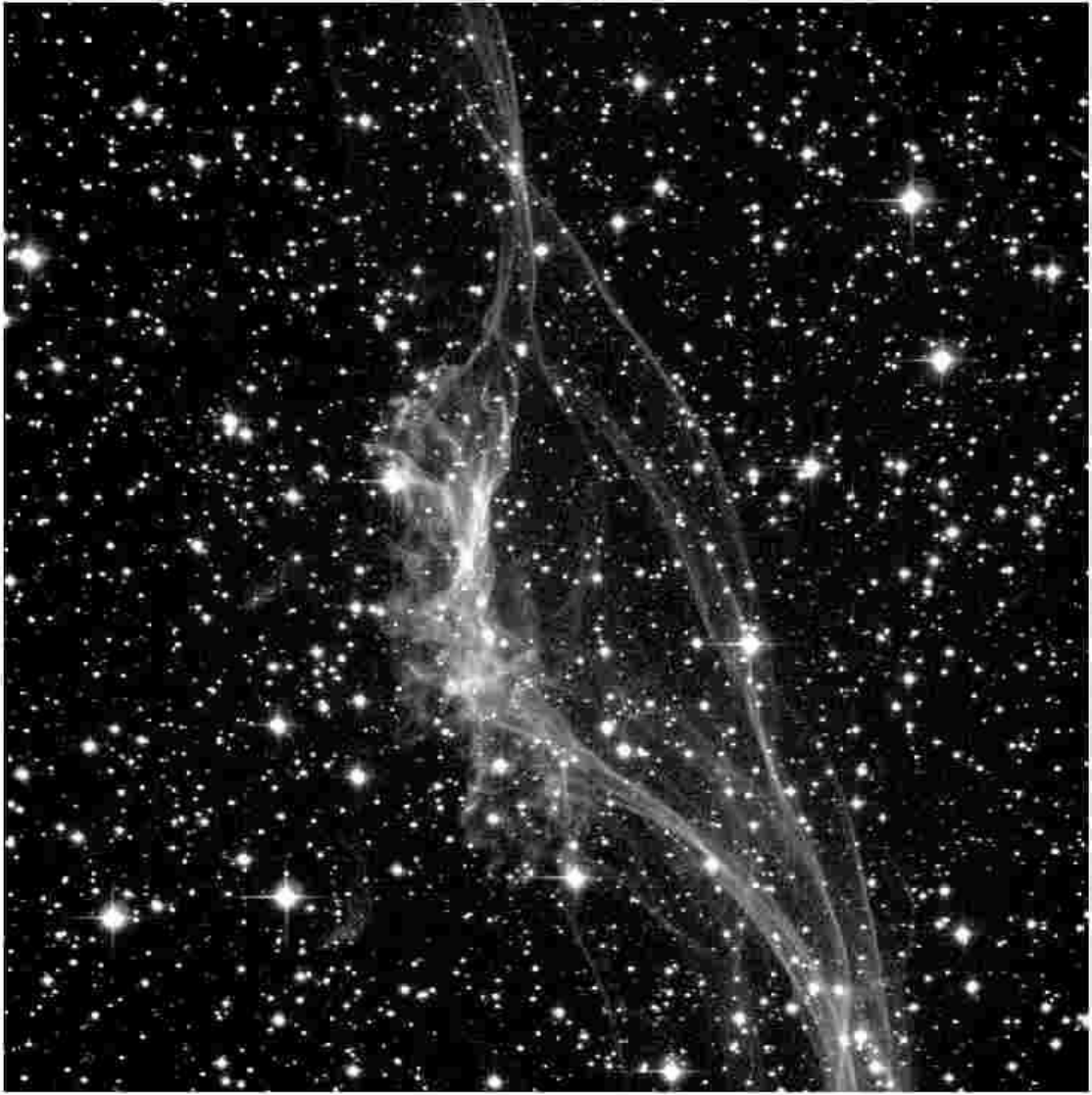}
\caption{MDM 1.3~m H$\alpha$ image of the southwest cloud
region. 
Image as shown covers 
a $11' \times 11'$ FOV. North is up and east is to the left.} 
\label{fig:mdm_ha_13} 
\end{figure}

The beautifully complex and delicate structure of the interaction 
of the remnant's shock front with this isolated 
ISM cloud is shown in the positive 
image of Figure~\ref{fig:mdm_ha_13}. 
Relatively undisturbed filaments 
can be seen extending north and south of the cloud, 
marking the current location of the remnant's advancing 
shock front toward the west. 
Similar appearing filaments lie west of the cloud and likely 
represent foreground and background portions of the blast wave 
unaffected by any cloud interaction. 

\begin{figure} 
\includegraphics[bb=20 20 575 455,width=6in,height=6in,keepaspectratio=true]{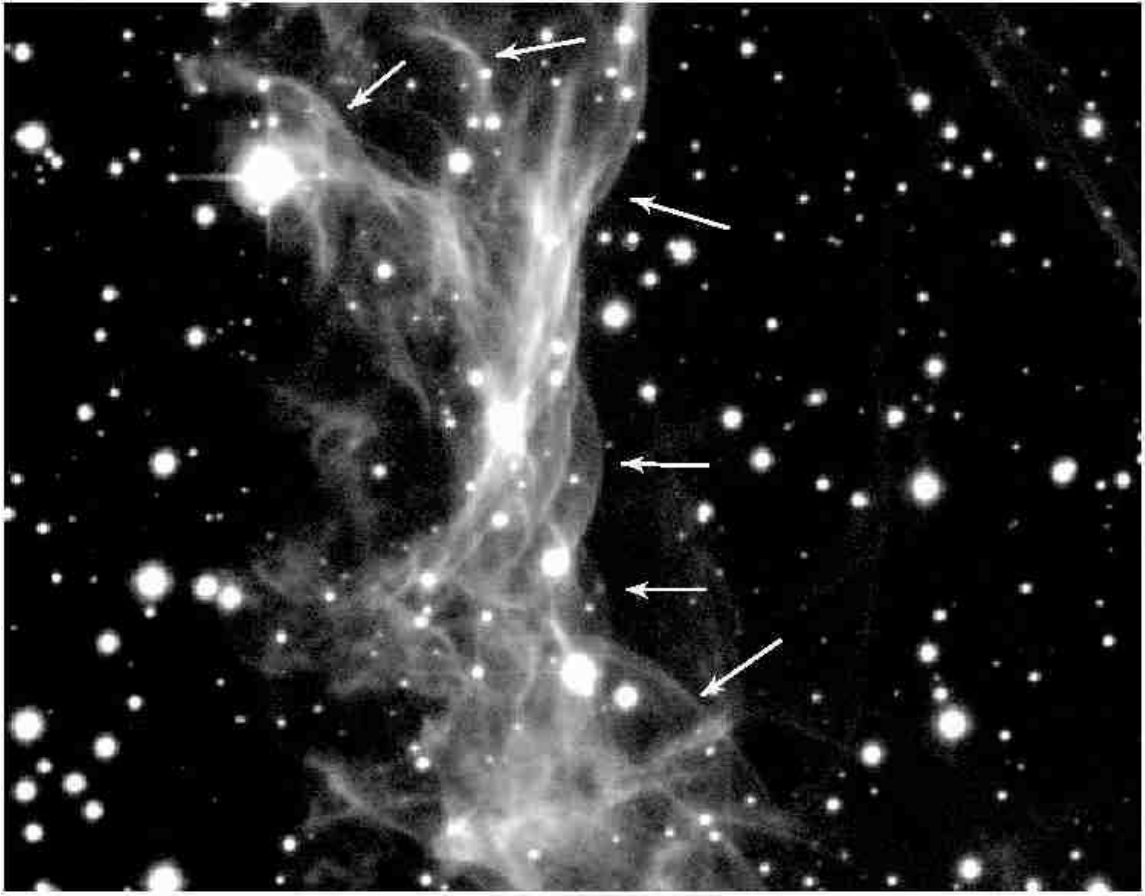}
\caption{MDM 2.4~m H$\alpha$ image of the southwest cloud.
Arrows mark 
the positions of several small shock fronts inside the cloud. 
Image as shown covers 
a $3.5' \times 2.7'$ FOV. North is up and east is to the left. The image has 
been scaled logarithmically to show the internal shocks.} 
\label{fig:ha24} 
\end{figure}

The positions of the filaments around the cloud strongly 
suggest that the cloud has only recently been impacted by the shock
wave. 
This conclusion is supported by the presence of numerous smaller
shock 
fronts visible 
within the cloud itself (Fig.~\ref{fig:ha24}). These ``shocklets'' 
consist of sharp, strong H$\alpha$ emission filaments. They 
reveal the progression of a seemingly fractured shock front 
traveling through the cloud's inhomogenous internal structure. 
These shocks also exhibit strong curvatures, often with different
shock 
propagation 
directions due to diffraction around density variations in the cloud. 
Such internal shocks have dimensions 
$\sim$ 10\arcsec~long, or roughly 7\% -- 10\% of the cloud diameter, and
their 
size and frequency provide clues to the internal density 
structure of this ISM cloud (see Section 4.1). 

The cloud's optical appearance is considerably different 
when viewed in other emission lines. 
Figure~\ref{fig:mdm_ha_o3_s2} compares the cloud as seen in 
H$\alpha$, [\ion{O}{3}], 
[\ion{S}{2}] and [\ion{O}{1}]. 
Whereas, the H$\alpha$ image shows a fairly sharp and filamentary 
appearance, the [\ion{O}{3}] image is largely diffuse, while 
that of [\ion{S}{2}] shows a more clumpy structure embedded in a 
diffuse component. Although the [\ion{O}{1}] $\lambda$6300 emission
is quite weak, its structure does not appear to closely resemble 
that of the other three line emission images. 

Interestingly, the cloud's [\ion{O}{3}] and [\ion{S}{2}]
morphologies 
are substantially different to those seen in most other regions of the 
Cygnus Loop for these emission lines \citep{fesen82,hester94}. 
Compared to most bright filamentary regions of the 
Cygnus Loop, which exhibit sharp and bright [\ion{O}{3}] filaments, 
the southwest cloud's [\ion{O}{3}] structure 
is strikingly diffuse having few readily identifiable filaments. 
In addition, while H$\alpha$ and [\ion{S}{2}] images are
similar looking 
for most other regions in the remnant, that does not hold true here. 
The cloud's appearance in [\ion{S}{2}] is quite clumpy, 
with a few bright knots present, particularly in the cloud's north and 
south ends. Also, the location of the peak [\ion{S}{2}] 
and H$\alpha$ emissions in the cloud do not coincide. 
The same situation appears to be true for [\ion{O}{3}] as well. 
Finally, the [\ion{O}{1}] image shows little correlation to the 
[\ion{S}{2}] image which might be expected in the presence 
of dense and lower ionization clumps. 

To better visualize the cloud's spatial emission differences, 
the H$\alpha$, [\ion{O}{3}] and [\ion{S}{2}] images 
were combined into a false color composite image 
(Fig.~\ref{fig:colorcomp}). In this composite image,
Balmer-dominated filaments 
appear red, strong [\ion{O}{3}] filaments appear blue, 
and strong [\ion{S}{2}] regions 
are shown as green. Regions with strong H$\alpha$ and [\ion{S}{2}] 
emission show up yellow, 
while regions strong in H$\alpha$ and [\ion{O}{3}] appear magenta.
Regions 
bright in [\ion{O}{3}] and [\ion{S}{2}] would look cyan, but this 
color is noticeably absent. Along the southern periphery, and to a 
lesser extent 
the northern periphery, H$\alpha$ emission is often seen immediately
alongside strong [\ion{O}{3}] emission. 
The eastern edges of the cloud are brightest in [\ion{S}{2}] emission 
where, in comparison, the H$\alpha$ is relatively faint. 

\subsection{Optical Spectrophotometry} 

To quantify comparisons of the line emission strengths in
[\ion{O}{3}], 
[\ion{S}{2}], and H$\alpha$, several apertures of various length 
($\sim$ 5\arcsec -- 20\arcsec) with a fixed 1.5\arcsec~ width were
extracted 
from the flux-calibrated images. The slices were made across the
three main 
types 
of emission filaments present: 
suspected Balmer-dominated filaments, radiative shocks, and areas
with 
unusually strong [\ion{S}{2}] and [\ion{O}{3}]. 

\subsubsection{Balmer-Dominated Filaments} 

\begin{figure} 
\includegraphics[bb=20 20 575 455,width=6in,height=6in,keepaspectratio=true]{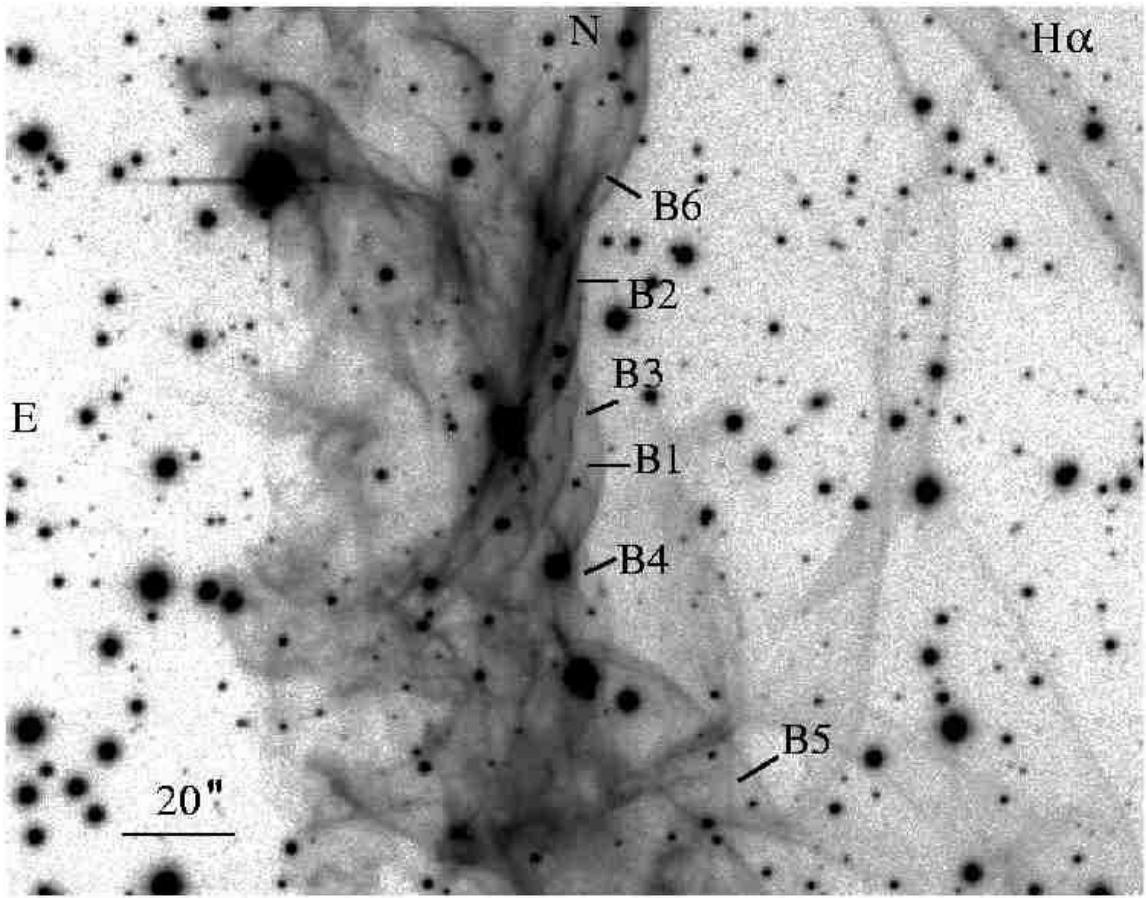} 
\caption{MDM 2.4 m H$\alpha$ image of the southwest cloud. Labels 
B1--B6 mark the location of several Balmer-dominated filaments discussed
in the text.} 
\label{fig:br} 
\end{figure}

\begin{figure} 
\includegraphics[bb=20 20 575 575,width=6in,height=6in,keepaspectratio=true]{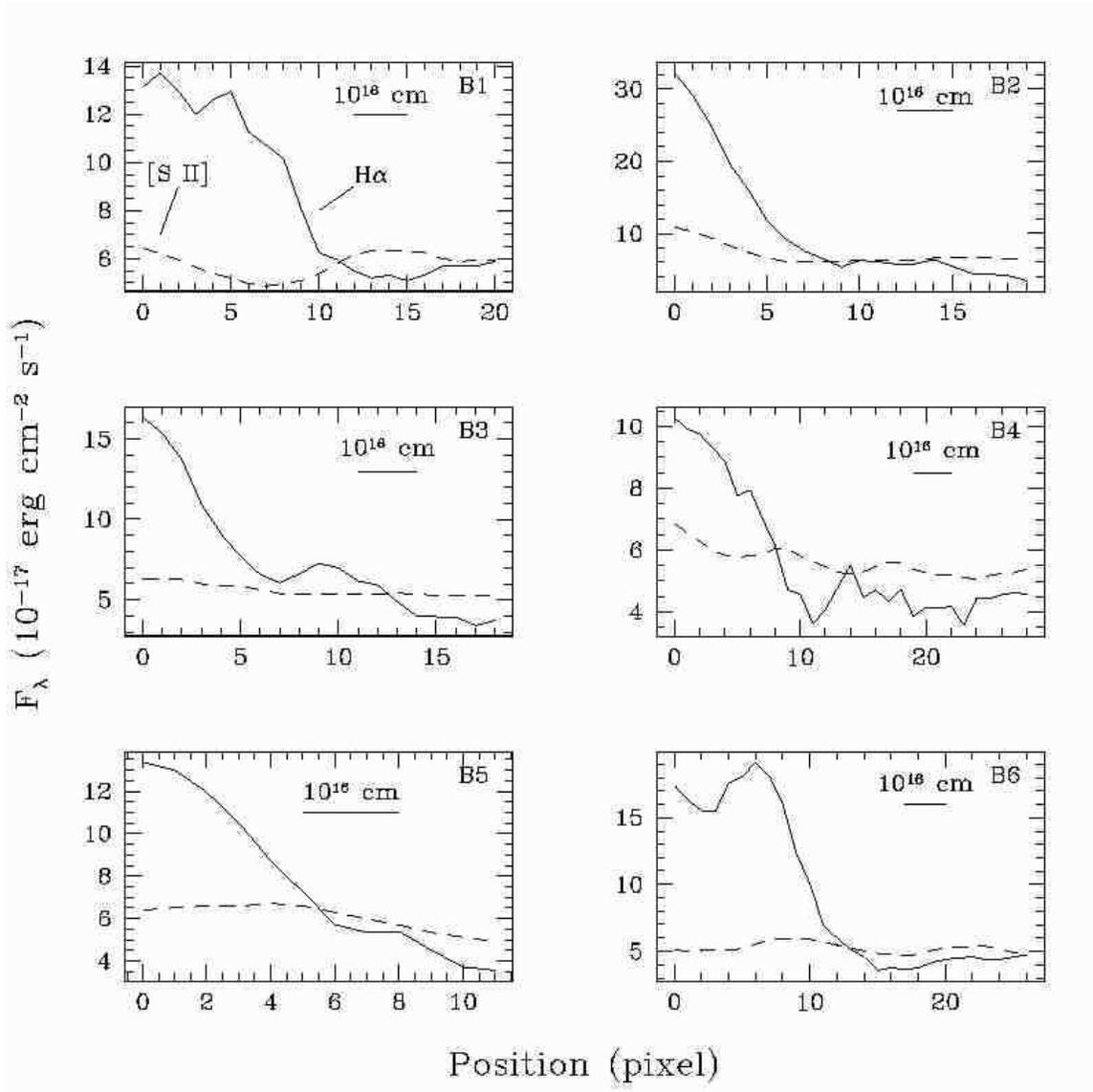}
\caption{One-dimensional profiles for Balmer-dominated shock
front regions 
B1 -- B6 labeled in Figure~\ref{fig:br}. Plots show line strength 
of H$\alpha$ (solid line) versus [\ion{S}{2}] $\lambda\lambda$6716,6731 
(dashed line). Measured fluxes are per 1.5\arcsec~$\times$ 0.5\arcsec~ 
pixels.} 
\label{fig:bro} 
\end{figure}

As already noted above in reference to Figure~\ref{fig:ha24}, numerous 
small scale shocks are present along the western edge of the cloud. For 
several of these suspected Balmer-dominated shocks, one-dimensional line 
flux profiles, measured normal to the shock front, were extracted from 
the H$\alpha$ and [\ion{S}{2}] images. The positions of these aperture 
extractions are shown in Figure~\ref{fig:br} and are labeled B1 -- B6. 
The resulting 1D cuts across the six regions are shown in 
Fig.~\ref{fig:bro}. The locations of our two long-slit spectra coincide 
with Positions B1 and B2. In all six regions, H$\alpha$ emission is seen 
to increase relative to [\ion{S}{2}]. The relative increase occurs over
a spatial scale of a few times 10$^{16}$ cm. Given that these data have been
background subtracted, the observed variations in the H$\alpha$/[\ion{S}{2}]
intensity 
ratio represent real, relative flux variations in the
H$\alpha$ to [\ion{S}{2}] ratio. The [\ion{S}{2}] emission
appears relatively constant from panel to panel, consistent with a diffuse,
background. Furthermore, the relative increase between 
H$\alpha$ and [\ion{S}{2}] is consistent from panel to panel, including
the two spectroscopic Positions B1 and B2, suggesting
that the ratios seen for Positions B3--B6 are real and not an artifact of our
background subtraction technique.

\begin{figure} 
\includegraphics[bb=20 20 575 575,width=6in,height=6in,keepaspectratio=true]{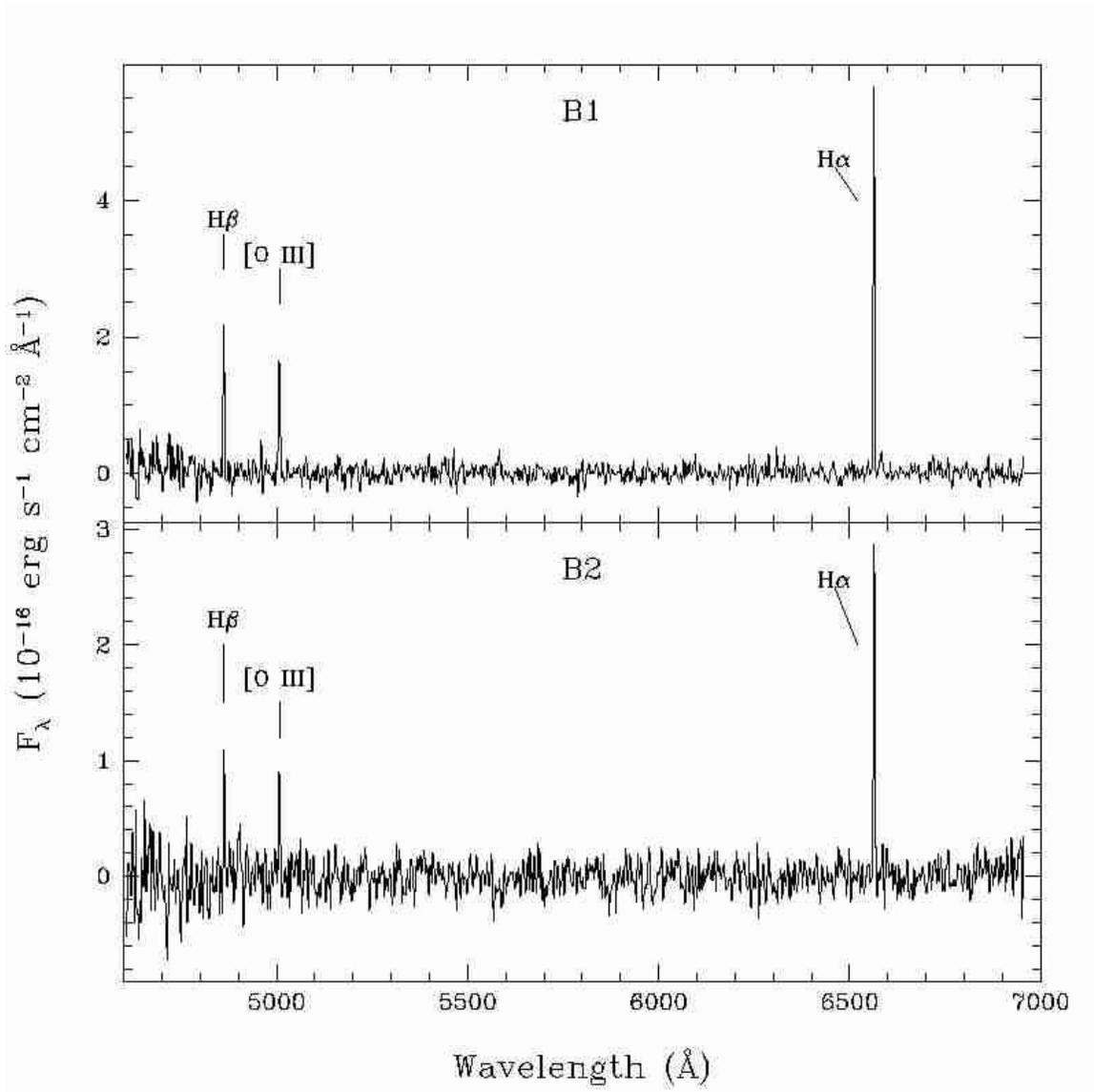}
\caption{Emission line spectra for Positions B1 and B2.} 
\label{fig:brs} 
\end{figure}

The long-slit spectra for Positions B1 and B2 (Fig.~\ref{fig:brs}) 
show bright Balmer emission, weak [\ion{O}{3}] emission, 
and no detectable [\ion{S}{2}] emission. The ratios of 
H$\alpha$/[\ion{O}{3}] for regions B1 and B2 are 3.2 and 4.3 respectively 
(see Table~\ref{tab:opticalregions}). These spectra 
are very similar to spectra found in other regions of the Cygnus Loop
located at or close to the shock front \citep{fesen82}. 
This similarity adds credence to the belief that the small shock fronts 
along the western edge of the cloud are Balmer-dominated filaments.

\subsubsection{[\ion{O}{3}]-Bright Regions} 

In several portions of the Cygnus Loop's brighter nebulosity, 
strong [\ion{O}{3}] emission has been seen adjacent to
Balmer-dominated 
emission. 
This has generally been interpreted as due to incomplete shocks 
\citep{blair91}. 
Incomplete shocks are often identified optically by 
[\ion{O}{3}]/H$\alpha$ ratios $\geq$ 3. 
In order to investigate whether incomplete shocks could explain 
the strong [\ion{O}{3}] emission along certain western portion of the
southwest cloud, 1D line flux profiles were extracted 
for several regions, marked O1 -- O4 in 
Figure~\ref{fig:reg}, from the [\ion{O}{3}] and H$\alpha$ images. 

The four regions selected all lie along the cloud's southern boundary 
where [\ion{O}{3}] emission is especially prominent (see color
image, 
Fig.~\ref{fig:colorcomp}). 
For all four regions we measured an [\ion{O}{3}]/H$\alpha$ ratio of
$2 - 3$. 
However, we suspect all four regions had background H$\alpha$ 
contamination from either Balmer-dominated filaments along the
cloud 
southern edge or projected diffuse, pre-shock gas. We estimate that 
this contamination could account for nearly half of the 
measured H$\alpha$ intensity. This would increase the observed 
[\ion{O}{3}]/H$\alpha$ ratio over 3, thereby indicating 
the presence of incomplete shocks. While this is likely to 
be true of the strong [\ion{O}{3}] regions O1 and O2, Positions 
O3 and O4 lie well behind the shock front making their 
[\ion{O}{3}] line emission nature less certain (see Section 4.2.2). 

\subsubsection{Radiative filaments} 

\begin{figure} 
\includegraphics[bb=20 20 575 575,width=6in,height=6in,keepaspectratio=true]{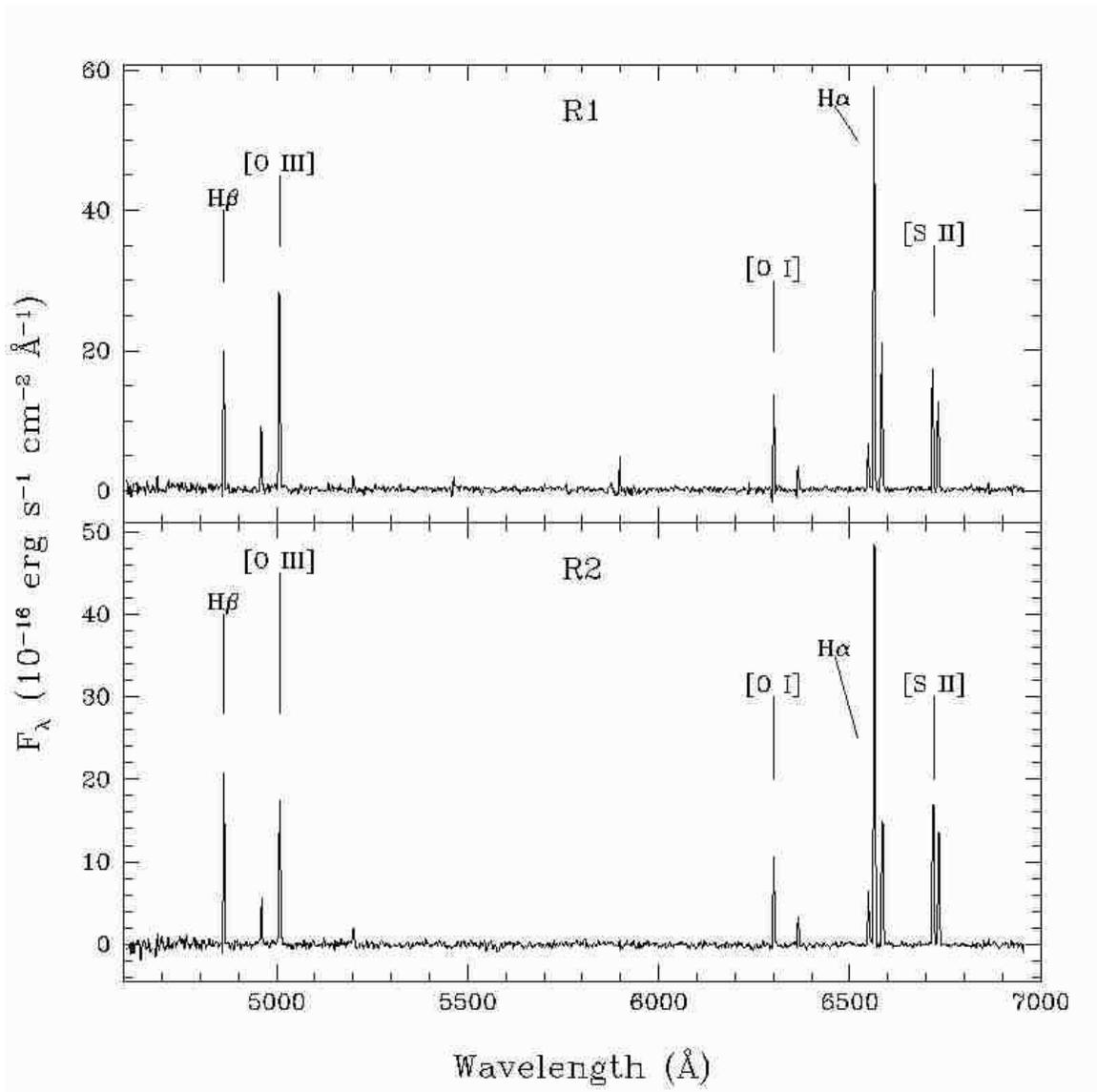} 
\caption{Emission line spectra for slit Positions R1 and R2.} 
\label{fig:radspec} 
\end{figure}

Bright [\ion{S}{2}], [\ion{O}{3}], and H$\alpha$ emission regions are 
seen to lie east (downstream) 
of the Balmer-dominated filaments discussed above. Two 
regions, R1 and R2, extracted from our longslit spectra, are located 
near the center (R1) and to the north (R2) of the cloud 
(Fig.~\ref{fig:reg}), several arcseconds behind the 
Balmer-dominated filaments. At R1 we see bright H$\alpha$, [\ion{S}{2}], 
[\ion{O}{1}], [\ion{O}{3}] and H$\beta$, as shown in 
Figure~\ref{fig:radspec}. Line fluxes for the most prominent lines are 
listed in Table~\ref{tab:opticalregions}. At this position, we measured 
the electron-density-sensitive ratio [\ion{S}{2}] 6717/6731 
$=$ 1.4$\pm$0.1, which implies an electron density of 
10 -- 100 cm$^{-3}$ at T = 10$^{4}$ K. Region R2 
(Fig.~\ref{fig:radspec}) is located $\sim$ 45$''$ north of R1 and just a 
few arcseconds east of Balmer-dominated Position B2. Here the ratio 
[\ion{S}{2}] 6717/6731 was again measured at 1.4 $\pm$ 0.1. 

\begin{figure} 
\includegraphics[bb=20 20 575 575,width=6in,height=6in,keepaspectratio=true]{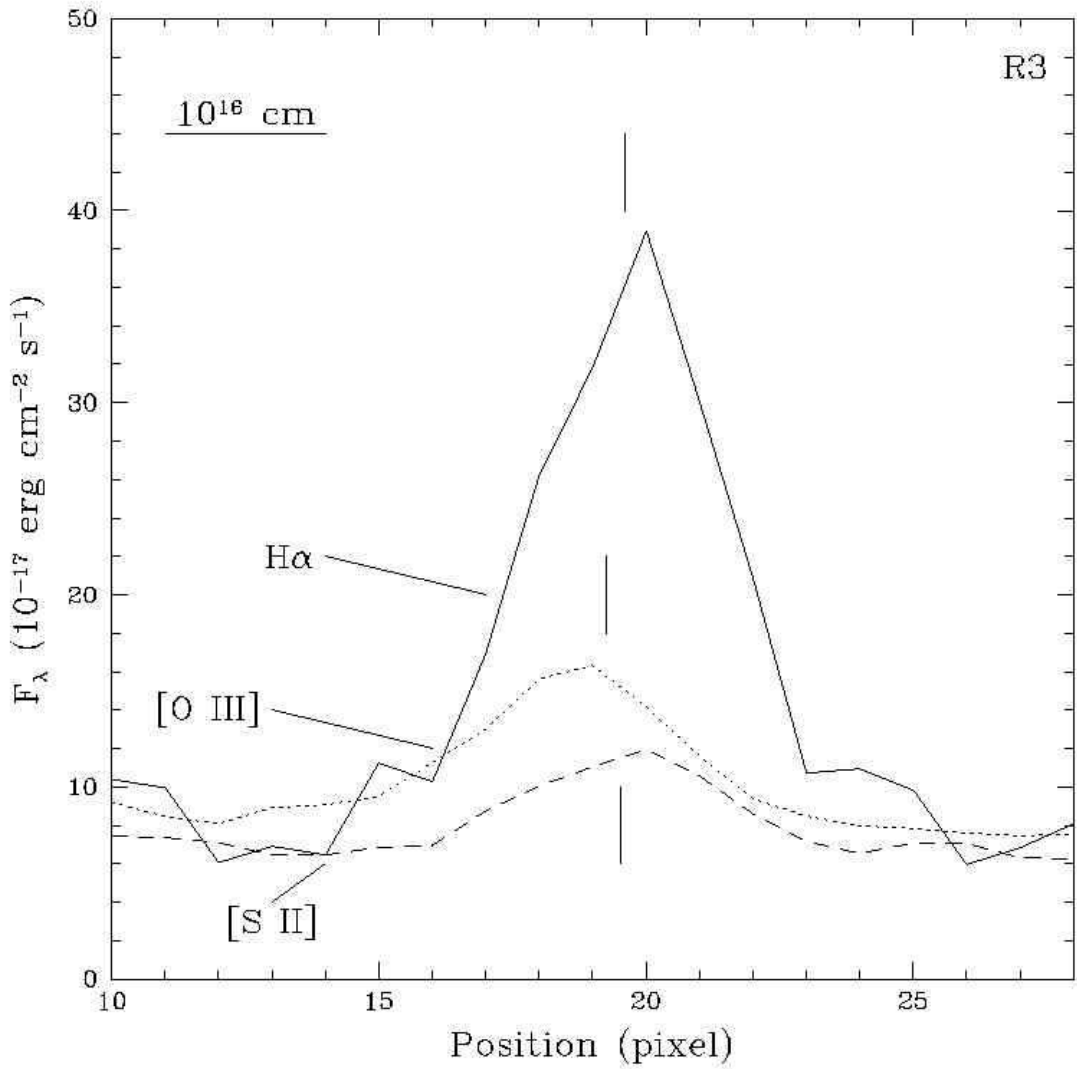}
\caption{One dimensional emission profile for the bright radiative 
filament Position R3. Measured fluxes are per 1.5\arcsec~$\times$
0.5\arcsec~ 
pixels.} 
\label{fig:regr3} 
\end{figure}

In addition to these 2D aperture spectra, we also 
extracted 1D emission profiles in H$\alpha$, 
[\ion{S}{2}], and [\ion{O}{3}] at Position R3, which is farther to the
north 
(see Fig.~\ref{fig:regr3}). The line intensity plot
(Fig.~\ref{fig:regr3}) 
shows maximum H$\alpha$ emission 
displaced some $< 5\times10^{15}$ cm ($\sim$ 2\arcsec) ahead of the 
[\ion{S}{2}] peak intensity 
and $\sim$ 10$^{16}$ cm ahead of the [\ion{O}{3}] emission peak. 

\subsubsection{Sulfur-Bright Regions} 

\begin{figure} 
\includegraphics[bb=20 20 575 455,width=6in,height=6in,keepaspectratio=true]{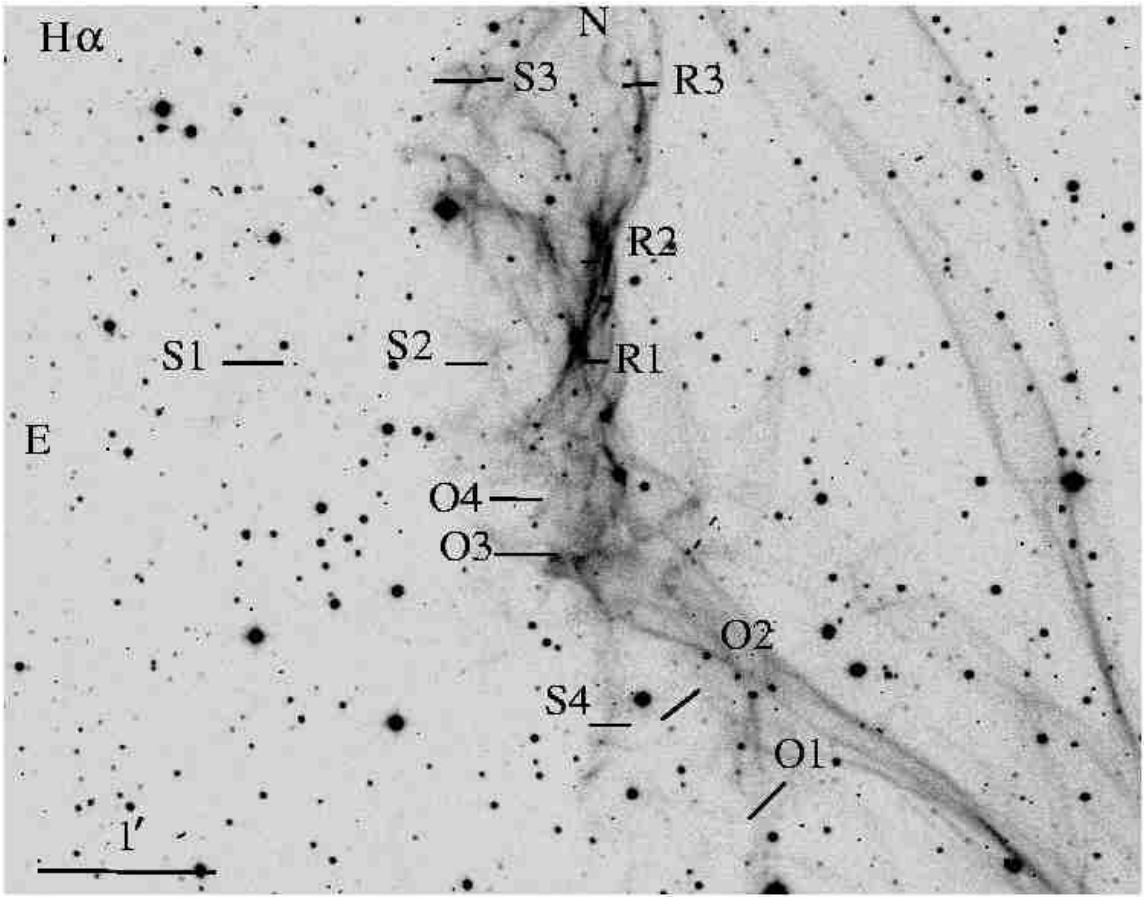}
\caption{MDM 1.3 m H$\alpha$ image of the southwest cloud with 
radiative, [\ion{O}{3}] bright, and [\ion{S}{2}] bright regions
marked 
where 1D line profiles were extracted.} 
\label{fig:reg} 
\end{figure}

\begin{figure} 
\includegraphics[bb=20 20 575 575,width=6in,height=6in,keepaspectratio=true]{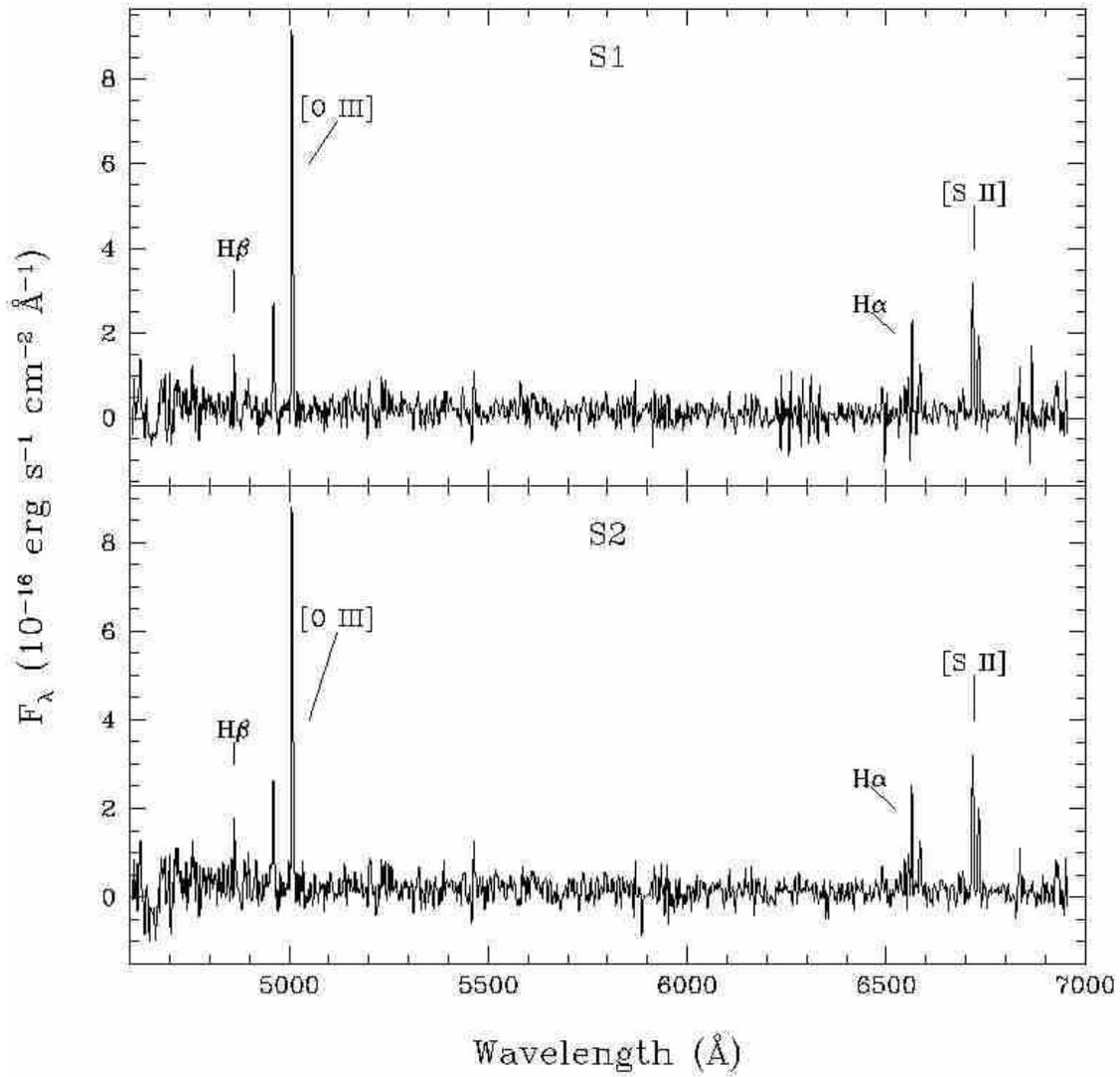}
\caption{Emission line spectra for the [\ion{S}{2}] bright slit 
positions S1 and S2.} 
\label{fig:s2spec} 
\end{figure}

There are several regions along the cloud's eastern edge 
which exhibit especially strong [\ion{S}{2}] emission. 
These can be seen as greenish clumps and filaments in 
Figure~\ref{fig:colorcomp}. 
Our central long-slit position crossed over two of these regions 
which we have labeled S1 and S2 in Figure~\ref{fig:reg}. 
Spectra extracted for these two regions 
are shown in Figure~\ref{fig:s2spec}. Although the spectra are fairly
noisy, 
they clearly show [\ion{S}{2}]/H$\alpha$ line ratios of $2 - 3$. 
This is much larger than the radiative filaments R1 and R2 values of 
0.53 and 0.59, 
and are unusual compared to other regions in the Cygnus Loop
\citep{fesen82}. 
The density-sensitive ratio in 
these two regions is $\approx$ 1.35, implying n$_e$ $_>\atop^{\sim}$ 
100 cm$^{-3}$ at T = 10$^{4}$ K), 
implying regions of somewhat higher 
density than those found in the radiative regions R1 and R2. 
These anomalously high [\ion{S}{2}]/H$\alpha$ knots can be
interpreted as 
clouds of gas that were shocked relatively long ago and have
recombined 
to the extent that the Balmer line emission from recombination has
begun 
to drop. The details depend upon the ratio of density to the radiation 
field, but this interpretation is supported by their location 
well behind the main shock. 

\begin{figure} 
\includegraphics[bb=20 20 575 575,width=6in,height=6in,keepaspectratio=true]{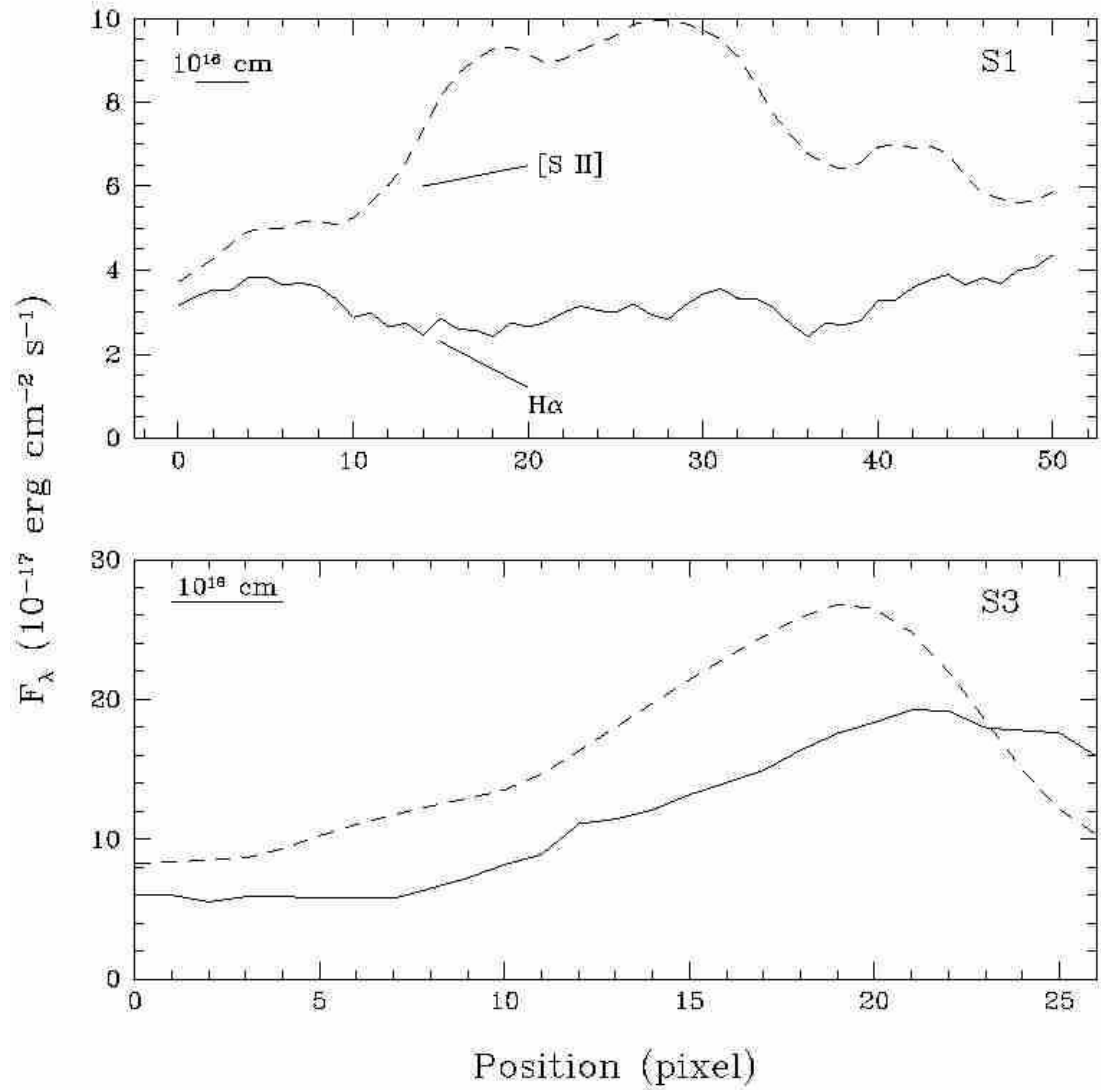}
\caption{One-dimensional profiles for the [\ion{S}{2}] bright
regions 
S1 and S3. Measured fluxes are per 1.5\arcsec~$\times$ 0.5\arcsec~ 
pixels.} 
\label{fig:brights2} 
\end{figure}

In addition to the S1 and S2 spectra, we extracted 
1D line profiles for two similarly greenish looking regions 
which we have labeled S3 and S4 on 
Figure~\ref{fig:reg}. 
Both show smaller [\ion{S}{2}]/H$\alpha$ ratios around 1.5. 
Furthermore, we find that these two regions differ from 
Regions S1 and S2 morphologically in that the [\ion{S}{2}] 
emission peak is displaced eastward from the peak H$\alpha$ 
emission by a $\sim$ 3 $\times$ 10$^{15}$ cm (see
Fig.~\ref{fig:brights2}). 
Here we see that in Region S1, the H$\alpha$ emission is relatively 
constant compared to the [\ion{S}{2}] emission, while in Region S3
we 
see that both the H$\alpha$ and [\ion{S}{2}] emission ramp up with
the 
[\ion{S}{2}] emission reaching a maximum intensity $\sim$
10$^{16}$ cm 
behind the H$\alpha$. 

\subsection{X-ray Imaging Spectroscopy} 
\label{sec:xrayimagspec}

As seen in Figure~\ref{fig:schmidtpspc}{\it b}, the global structure of 
the X-ray emitting gas in the Cygnus Loop closely resembles that of the 
optically-emitting gas. Furthermore, the X-ray morphology of the 
southwest cloud (Fig.~\ref{fig:schmidtpspc} bottom right), appears very 
similar to the global morphology of the southwest cloud seen in 
H$\alpha$. In the X-ray, we see an undisturbed shock front both north 
and south of the cloud. This shock front is coincident with the 
Balmer-dominated shock front seen in the KPNO image. Also, the bright 
X-ray emission seen south of the cloud seems to correspond to the more 
tangled Balmer emission seen south of the cloud. 

\begin{figure} 
\includegraphics[bb=20 20 575 455,width=6in,height=6in,keepaspectratio=true]{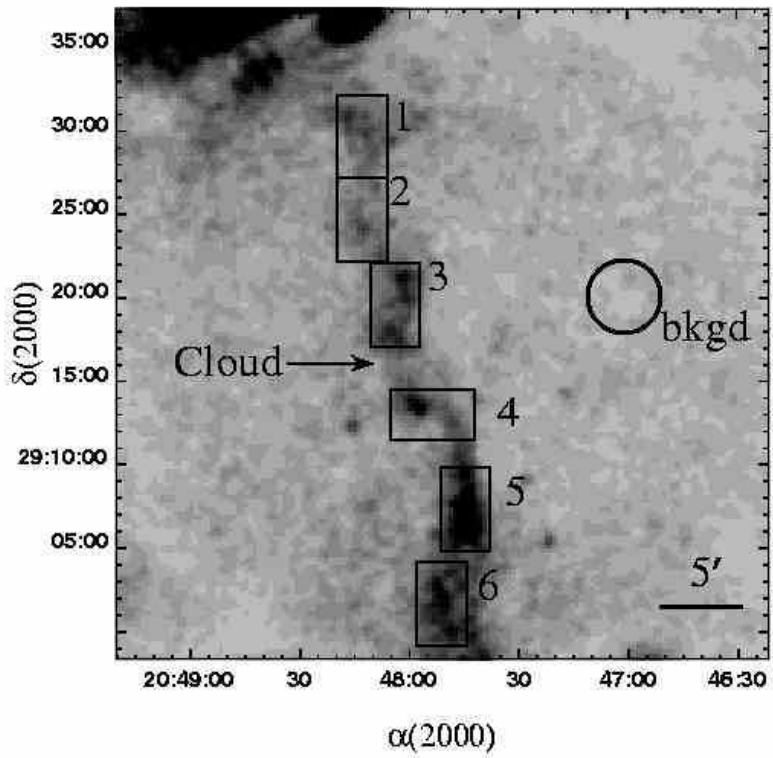}
\caption{{\it ROSAT} PSPC image of the southwest cloud. The image has
been smoothed to 1\arcmin~ resolution.} 
\label{fig:pspc} 
\end{figure}

Perhaps the most striking feature in the X-ray image is the absence of 
any appreciable emission from the cloud itself. In 
Figure~\ref{fig:pspc}, we have focused in on the cloud and surrounding 
shock front. There is bright X-ray emission, seen as spots, 
immediately north and south 
of the cloud, but the X-ray emission from the cloud itself is 
noticeably fainter than the surroundings. These bright X-ray spots 
are most likely due to enhanced emission as a result of a 
slight increase in density 
encountered by the X-ray emitting shock along the cloud's outer boundaries. 
Much slower shocks ($\leq$ 150 km s$^{-1}$)
would account for the lack of detectable X-ray emitting gas for the
main body of the cloud.

We label in Figure~\ref{fig:pspc} several regions around the southwest 
cloud where we extracted spectra for further analysis. The results of 
our fits are listed in Table~\ref{tab:regions}. Regions 1 and 2 are 
associated with the undisturbed shock front north of the cloud. Here the 
shock is progressing rather uniformly through the ISM, which is apparent 
by the smoothness of the shock front, as well as uniformity in X-ray gas 
temperature ($\sim$ 0.13--0.12 keV). Region 3, immediately north of the 
cloud, has a lower temperature (0.11 $\pm$ 0.03 keV), but higher flux 
(F$_x$ = 0.7 $\times$ 10$^{-12}$ erg cm$^{-2}$ s$^{-1}$) and emission 
measure (EM = 0.7 cm$^{-6}$ pc). A spatial examination of Region 3 
reveals that the bulk of the X-ray emission arises from two knots within 
the 5\arcmin$\times$3\arcmin~box. 

Region 4, immediately south of the cloud, is composed of the bright knot 
directly south of the cloud plus more diffuse X-ray emission trailing off 
to the west along the forward shock front. Here we again find the X-ray 
gas temperature to be $\sim$ 0.11 keV, which corresponds to a X-ray 
shock velocity of 290 km s$^{-1}$. However we find the X-ray flux and 
emission measure to be lower than in Region 3. Regions 5 and 6 are by 
far the brightest regions (10$^{-12}$ erg cm$^{-2}$ s$^{-1}$). As noted, 
these two regions are associated with bright Balmer emission seen south 
of the cloud in Figure~\ref{fig:schmidtpspc}. Here the shock front seems 
to be running into another ISM cloud, though the structure appears much 
more complicated than that of the southwest cloud. 

In general, our results show hints of a systematic temperature gradient 
relative to the cloud. The X-ray emitting gas temperature drops at the 
clouds northern edge when approaching the cloud from the north (Regions 
1, 2, \& 3). This pattern is then reversed south of the cloud, with the 
temperature of Region 4 being essentially identical to that of Region 3, 
and then increasing towards Regions 5 and 6. 

\section{DISCUSSION} 
\label{sect:disc} 

\subsection{Dynamics} 

Shock-cloud interactions can be characterized by four stages of evolution 
\citep{klein94}. The first stage occurs when the blast wave initially 
encounters the cloud, driving a strong shock into its face and forming a 
standing bow shock downstream. The second stage is cloud compression as the 
blast wave wraps around the rear of the cloud and re-converges upstream. A 
reflected shock forms at the re-converged shock front apex, which then 
travels back downstream into the cloud. This newly-formed shock, acting in 
concert with the slow internal cloud shocks traveling forward through the 
cloud, further compresses the cloud. The third stage involves a re-expansion 
of the cloud. This begins when the main cloud-shock reaches the downstream 
end of the cloud, causing a strong rarefaction to be driven back into the 
cloud, and leading to cloud re-expansion in the upstream direction. Finally, 
the cloud is destroyed as instabilities and differential forces due to the 
flow of intercloud gas past the cloud and cause it to fragment. 

A useful parameter for measuring a shocked cloud's evolution is what Klein 
et al. (1994) refer to as a ``cloud crushing time'', t$_{cc}$. The cloud 
crushing time is the characteristic timescale for a cloud to be crushed by a 
shock moving through the cloud, and is therefore wholly dependent upon the 
cloud shock velocity ($t_{cc} = r_c / v_s$). However, the cloud shock 
velocity v$_s$ is dependent upon the blast wave velocity v$_b$ (Klein et al. 
1994), so the cloud crushing time can be characterized by the blast wave 
velocity ($t_{cc} = \chi^{1/2}r_c/v_b$), where $\chi$ is the density contrast 
$\rho_{cloud}/\rho_{amb}$. In the case of the southwest cloud, we estimate 
t$_{cc}$ to be $\approx$ 1500 yr assuming an initial cloud radius of $0.2$ 
pc (from optical measurements), a blast wave velocity of 290 km s$^{-1}$, and 
a density contrast, $\chi$, of 5 \citep{ku84}. 

In their study, Klein et al. (1994) focused their analysis to clouds which 
are small compared to the blast wave. Small clouds are defined as those 
where the cloud crushing time $t_{cc}$ is of order the pressure variation 
time-scale $t_P$. The pressure variation time-scale for a dense cloud in a 
Sedov-Taylor blast wave is $t_P \simeq 0.1 r_c/v_b$. Taking a blast wave 
velocity of 290 km s$^{-1}$ for the southwest cloud 
(see Section~\ref{sec:xrayimagspec}) and assuming that the cloud 
was initially roughly 
spherical ($r_c \simeq 0.2$ pc), then the pressure time-scale is $\sim$ 
70 yr. 
Klein et al. (1994) define medium clouds 
as those with $t_{cc} {_>\atop^{\sim}} t_P$, so one can therefore view the 
southwest cloud in that context. 

\subsubsection{The Age of the Shock--Southwest Cloud
Interaction} 

The structure of the southwest cloud itself and associated neighboring shock 
fronts suggest an early stage of shock-cloud interaction, probably somewhere 
in between stages one and two described above, with an age of $\sim$ 1200 yr. 
Several pieces of evidence support this. 

First, Klein et al. (1994) argue that the wrap-around shocks will re-converge 
in a time of order 1.2 $t_{cc}$. Given the lack of a clear re-convergence 
point upstream of the southwest cloud, the cloud-shock interaction time must 
then be less than the time required for the wrap-around shocks to reconverge 
upstream of the cloud ($\approx$ 1800 yr). This conclusion is supported by 
the presence of numerous shock fronts inside the cloud lying at distances of 
up to 2\arcmin--3\arcmin~to the east (behind) the line of undisturbed 
H$\alpha$ shock front filaments. This shock structure is what would be 
expected at a relatively early shock--cloud interaction phase. 

Secondly, shock--cloud models predict the formation of vortices along cloud 
edges as the shock passes by them. These vortices, and the resulting 
turbulence they create, are the result of Kelvin-Helmholtz (K-H) 
instabilities that form at the cloud-shock interface tangential to the shock 
normal. The formation of K-H instabilities occurs on a timescale comparable 
to the cloud crushing timescale. In Figure~\ref{fig:mdm_ha_13}, several 
finger-like structures are seen to the north and south of the cloud. At 
first glance, these structures look like the result of K-H shearing. However, 
instabilities resulting from K-H shearing tend to follow the flow of the 
shock, whereas these fingers are flowing in the opposite sense. Therefore, our 
optical imaging of the southwest cloud shows no hint of K-H vortices along 
the edges of the cloud, thereby also suggesting an age less than $\sim$ 
1500 yr. 

Both age estimates above agree with a simple, crude age estimate of the 
cloud-shock interaction derived from assuming a roughly spherical cloud of 
radius 0.2 pc and shock velocity of 290 km s$^{-1}$. It would take $\sim$ 
1500 yr for the shock to advance from front to the rear of the cloud, 
consistent with the location of the undisturbed shock front position as 
viewed in the H$\alpha$ images. 

A final piece of evidence involves the lack of a clearly detectable standing 
bow shock in the postshock gas. Klein et al. (1994) state that a standing bow 
shock will form in a time $t_b = 2r_c/v_b$. Using the above numbers as 
estimates, we find that the bow shock will form in 1300 yr. The lack of such 
a structure visible in the {\it ROSAT} X-ray image can be taken as evidence 
that this interaction is therefore much younger than a cloud crushing time, 
and at least as young as the $\sim$ 1300 yr bow shock formation timescale. 

Therefore, given the assumed size of the cloud and shock velocity as measured 
from the X-ray observations, we suggest that the age of this cloud--shock
interaction is $\approx$ 1200 yr. We estimate an uncertainty of $\pm$ 
500 yr based
upon the fact that we do not know the initial size of the cloud, and the fact
that the shock velocity is an estimate based on the postshock gas 
temperature.

\subsubsection{Cloud Shock Structures} 

As seen in Figures~\ref{fig:mdm_ha_13} and \ref{fig:ha24}, the cloud contains
 a rich variety of shock structure on several scale lengths. 
As predicted by Klein et al. (1994), the main blast-wave wraps around 
the cloud with a curvature of order one cloud diameter. Of particular 
interest is the presence of several smaller shocks seen within the cloud 
itself. Intracloud shocks permeate the cloud on various scales, and at 
various angles to the main shock direction. This is suggestive of shock wave 
diffraction, analogous to a water wave passing through a narrow channel and 
exiting on the other side with curvature affected by
 the channel width. Similar 
arguments can be made with respect to the southwest cloud. 

Initially, the cloud might have consisted of a low density intracloud medium 
permeated by small scale density variations (clumps). As the main shock wave 
progresses through the cloud, it is slowed by these density enhancements, 
while the intracloud shock travels in between them. These density 
enhancements go through the same cloud-shock interactions, only on a much 
smaller scale. In some cases the shock might stall, given that it is already 
traveling at a much slower velocity. However, the undisturbed shock will 
continue to carry momentum and energy through the less dense regions in the 
cloud, resulting in shocks with curvatures of order the clump separation. 

The shock fronts within the cloud are about 7\arcsec~--10\arcsec~long. If the 
shock ends are ``attached'' to small regions of higher density, then the 
average distance between higher density clumps in the cloud is 
$_<\atop^{\sim}$10\arcsec. Assuming that the cloud was initially 
spherical with an angular size of $\sim$ 3\arcmin, we estimate that 
density fluctuations within this ISM cloud make up as much 
as $\sim$ 20\% of the volume of the cloud. Furthermore, in order for this 
type of shock diffraction to occur, the density fluctuations must be at least 
of order of the density contrast between the cloud and the ISM. 

Models for shock-cloud interactions often use a homogeneous initial density 
with a well defined boundary between the cloud and the intercloud zones. 
However, both our optical and X-ray observations show clearly that this is 
not the case for the southwest cloud. Decreased densities along a cloud's 
outermost portions should result in a lower postshock density, and therefore 
a larger postshock cooling zone. This situation is actually seen along the 
cloud's southern edge, where there is a large band of diffuse [\ion{O}{3}] 
emission. Similarly, but less dramatically, the same appears to be occurring 
in the north. 

\begin{figure} 
\includegraphics[bb=20 20 575 455,width=6in,height=6in,keepaspectratio=true]{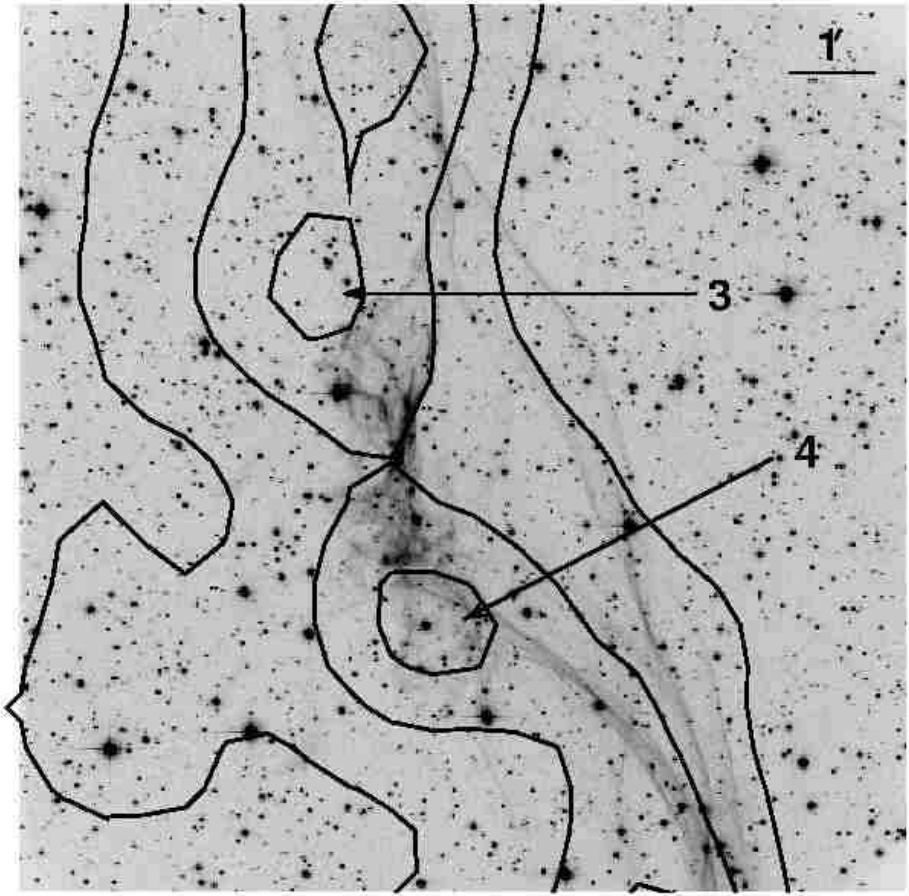}
\caption{H$\alpha$ image of the southwest cloud overlaid with {\it
ROSAT} 
PSPC X-ray 
contours. Contour intervals shown are 10.7, 12.8, and 14.9 counts.
Regions 3 
and 4 correspond to two bright X-ray knots visible in 
Figure~\ref{fig:schmidtpspc} and tabulated in
Table~\ref{tab:regions}.} 
\label{fig:ha13-xray} 
\end{figure}

We find support for a gradual drop in density by noting that the X-ray 
emission is brightest along the north and south of the cloud, and centered 
somewhat east of the main shock front, as marked by Balmer-dominated 
filaments, that has already passed the cloud (see Fig.~\ref{fig:ha13-xray}). 
If the edges of the cloud are of a lower density than the cloud itself, then 
the shock will not be slowed as much in its passage through this area. 
The comparison of the X-ray contours to the H$\alpha$ emission show little 
evidence for optical line emission near the X-ray knots, and this is 
confirmed by the lack of emission in [\ion{O}{3}] and [\ion{S}{2}].

\subsubsection{Comparisons to the Southeast Cloud} 

Compared to the Cygnus Loop's southeast cloud \citep{fesen92,graham95,
levenson01}, the southwest cloud appears to be in a much younger stage of 
shock-cloud interaction. The southeast cloud was initially identified as a 
small cloud in the late stage of shock-cloud interaction \citep{fesen92}. 
Indeed, the resemblance to late-stage numerical models of shock-cloud 
interactions is striking \citep{bedogni90, stone92}. However, more recent 
X-ray and optical studies \citep{graham95, levenson01} suggest that the 
shock wave is interacting with the tip of a much larger cloud. Either way, 
however, the southwest and southeast clouds exhibit several similarities. 
Balmer-dominated filaments are seen to trace out the shock fronts as they 
wrap around and attempt to engulf both clouds. In the southwest cloud, X-ray 
hotspots are evident north and south of the cloud, while in the southeast 
cloud there is a similar hotspot along the southern edge of the cloud. Such 
hotspots seem to occur where the undisturbed shock front connects to the 
engulfing shock front. Finally, radiative filaments are seen to trail behind 
the Balmer-dominated filaments for both clouds. 

However, there are also several distinct features of the southwest cloud 
which set it apart from the southeast cloud due to its youthful shock-cloud 
interaction. In the southeast cloud, there is bright X-ray emission 
associated with the shock as it wraps around the cloud. In contrast, we find 
virtually no X-ray emission associated with the main part of the southwest 
cloud. Also, in the southeast cloud a reverse shock is seen in the X-ray 
\citep{graham95}. No such reverse shock appears to exist in the southwest 
cloud. Given that a reverse shock will form shortly after the shock hits the 
cloud, the lack of a reverse shock suggests that it has not had time to form. 
A standing bow shock will only be seen if there is enough swept up material. 
The southwest cloud may be of a low enough density so that a standing X-ray 
bow shock will not form since the swept up gas density will never be above 
the critical density required for efficient heating and subsequent cooling. 

\subsection{Radiative Properties} 

\subsubsection{Optical} 

In H$\alpha$ images, we see undisturbed Balmer-dominated filaments tracing out 
the shock front north and south of the cloud. Such filaments also mark where 
the shock front has progressed around the back side of the cloud. We also 
see Balmer-dominated emission coming from several small shocks inside the 
cloud.

\begin{figure} 
\includegraphics[bb=20 20 575 552,width=6in,height=6in,keepaspectratio=true]{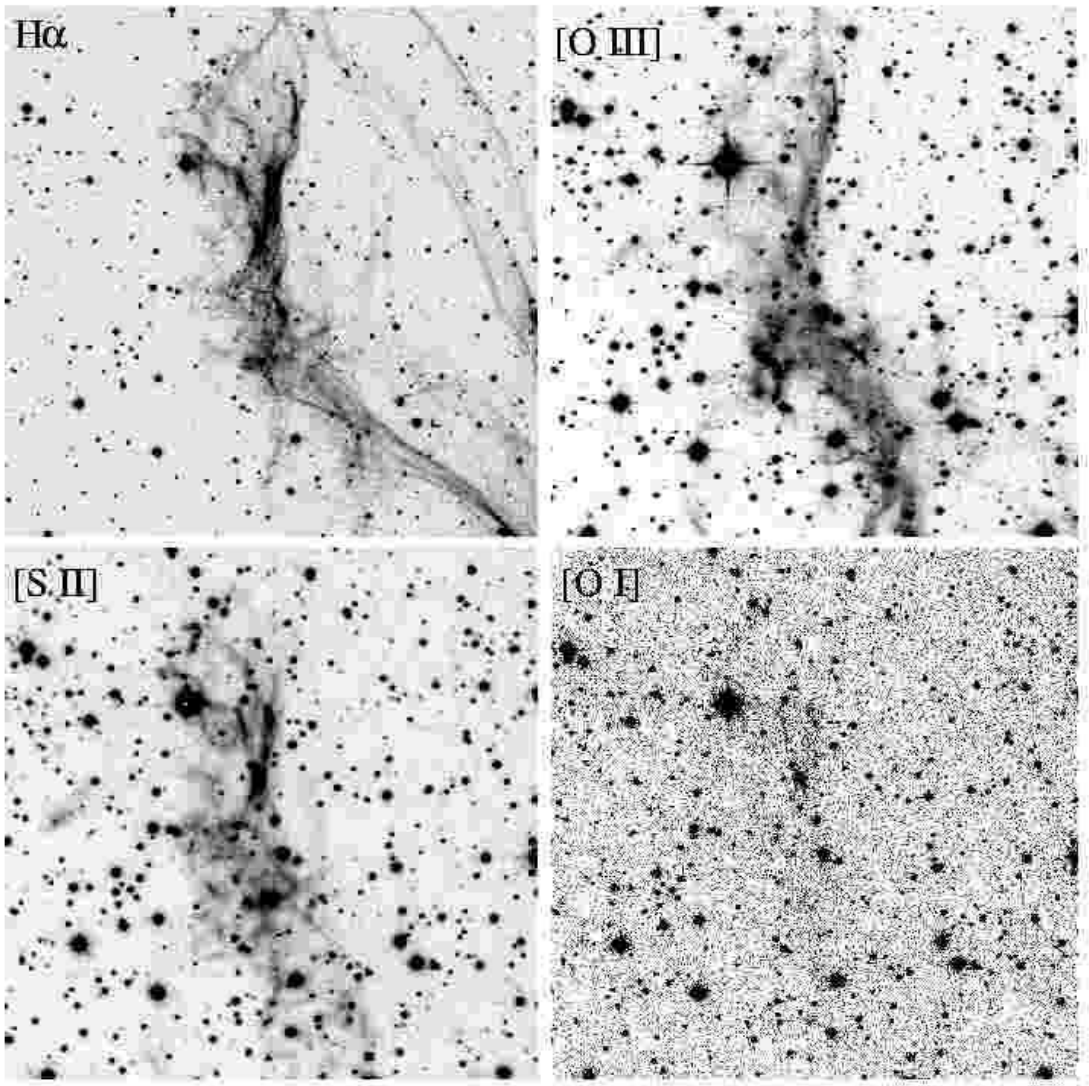}
\caption{Images of the southwest cloud in H$\alpha$, 
[\ion{O}{3}], [\ion{S}{2}], and [\ion{O}{1}]. Each panel covers a
region 
of $7.7' \times 7.7'$. North is up and east is to the left.} 
\label{fig:mdm_ha_o3_s2} 
\end{figure}

The cloud's optical appearance changes radically when viewed in line 
emissions from conventional radiative-type emission regions. While the H
$\alpha$ emission is characterized by thin, bright filamentary structure, 
just the opposite is true for its [\ion{O}{3}] emission. As shown in 
Figure~\ref{fig:mdm_ha_o3_s2}, the [\ion{O}{3}] emission is largely diffuse, 
with just a few sharp filaments. In addition, there is the relative lack of 
[\ion{O}{3}] emission north of the cloud compared to the south. A similar 
asymmetry is also seen in H$\alpha$ which might imply a density gradient 
along the north-south axis of the cloud. In contrast to both the [\ion{O}{3}] 
and H$\alpha$ emission morphologies, the cloud's strong [\ion{S}{2}] emission 
structure is limited to a few sharp filaments and some faint, diffuse 
emission patches (Fig.~\ref{fig:mdm_ha_o3_s2}). 

\begin{figure} 
\includegraphics[bb=20 20 575 524,width=6in,height=6in,keepaspectratio=true]{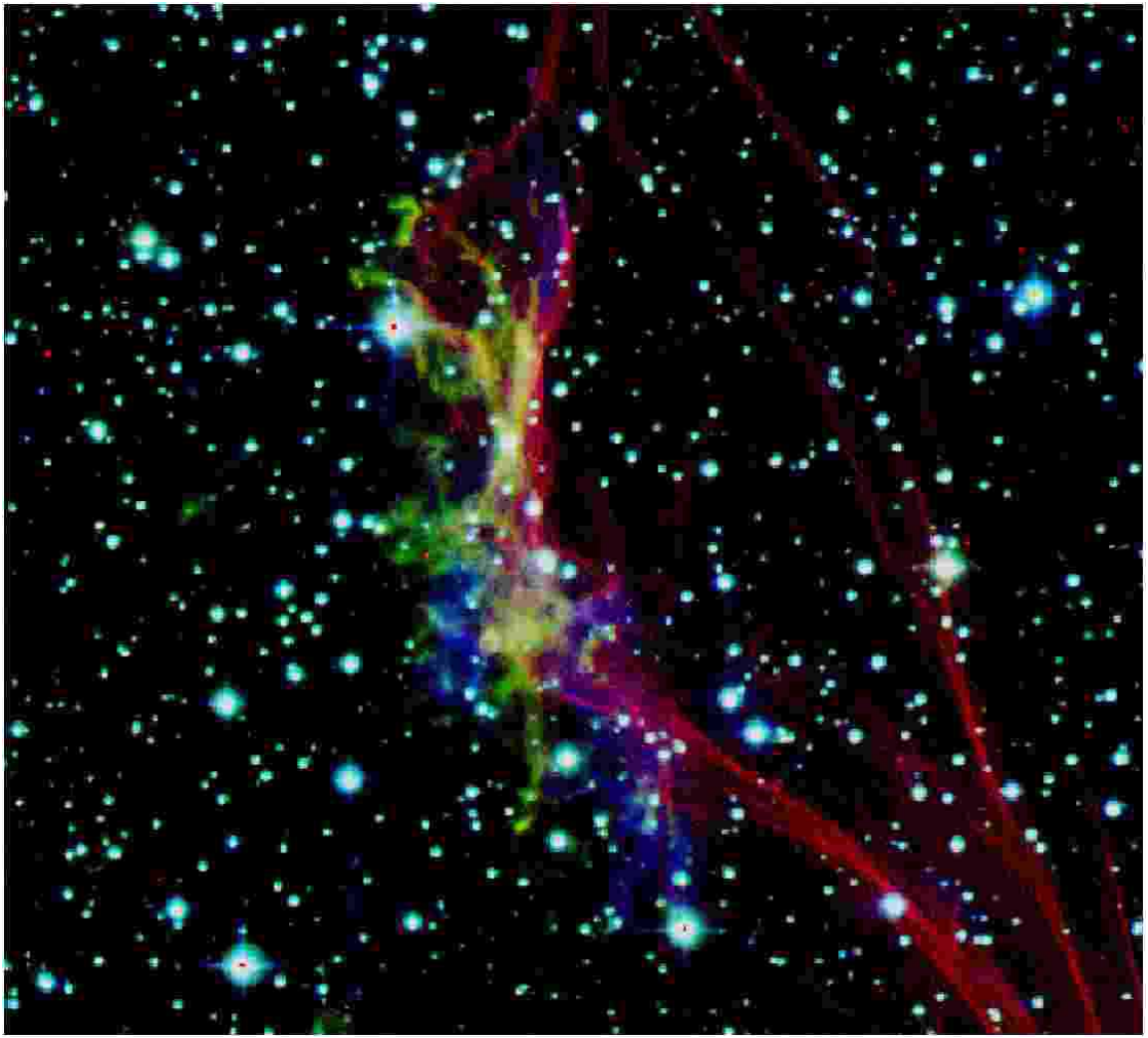}
\caption{False color composite image of the southwest cloud. 
H$\alpha$ is in red, 
[\ion{O}{3}] in blue, and [\ion{S}{2}] in green.} 
\label{fig:colorcomp} 
\end{figure}

Compared to the remnant's bright radiative northwestern and northeastern limb 
filaments (e.g. \citet{hester87}), the southwest cloud shows both 
similarities and differences when the H$\alpha$, [\ion{S}{2}], and 
[\ion{O}{3}] emissions are viewed together (Fig.~\ref{fig:colorcomp}). 
The cloud's northern extremity shows postshock line emission stratification 
like that discussed by \citep{hester87}. That is, Balmer-dominated filaments 
are followed closely by a region strong in [\ion{O}{3}] emission followed in 
turn by strong [\ion{S}{2}] line emissions (Fig.~\ref{fig:regr3}). Also, in 
the eastern part of this northern section, one finds bright, well defined 
[\ion{S}{2}] filaments, some diffuse H$\alpha$ emission but no corresponding 
[\ion{O}{3}] emission, suggesting postshock cooling regions farther 
downstream ($\sim$ 10$^{17}$ cm) from the shock front. 

In general, one finds Balmer-dominated filaments that may or may not be 
followed by [\ion{S}{2}] or [\ion{O}{3}] radiative emission. This is 
consistent with a resolved postshock cooling zone where first strong 
[\ion{O}{3}] line emission is seen followed by [\ion{S}{2}] and H$\alpha$. 
In the possibly older, denser shocked cloud regions only [\ion{S}{2}] 
emission remains. The lower density regions of the cloud will have longer 
postshock cooling times which may account for the diffuse nature of the 
cloud's [\ion{O}{3}] emission structure. The lack of clear line 
stratifications in the cloud's southern half may be an indication of a 
generally lower density compared to the north. This, in turn, would have led 
to the presence of numerous small scale shock structures like that seen in 
H$\alpha$. 

Region S1 is positioned well behind ($\sim$ 1\arcmin) the advancing shock 
front. As shown in Figure~\ref{fig:s2spec} and Figure~\ref{fig:brights2}, 
there is little Balmer emission, but bright emission from forbidden lines. 
The lack of Balmer emission, coupled with the strong emission in [\ion{S}{2}] 
and [\ion{O}{3}] suggest that this is a region of strong radiative cooling. 
Furthermore, in Figure~\ref{fig:colorcomp}, we find that the sulfur-bright 
regions are more compact than the diffuse oxygen-bright regions. This is 
consistent with our assumption that the [\ion{S}{2}] emission is arising from 
regions of cooler, higher density gas, which might be enveloped in a warmer 
[\ion{O}{3}] bright shell. 

In conclusion, while the cloud's overall shock structure is not unlike that 
expected for a recently shocked cloud, the structure and complexity of the 
line emission features was somewhat unexpected. Like elsewhere 
in the Cygnus Loop, sharp Balmer-dominated filaments nicely trace out the 
shock front. But the emission structure seen in other emission bands differs 
sharply from that seen in other regions of the remnant \citep{levenson98}. 
This is especially true in terms of the largely diffuse [\ion{O}{3}] emission 
and the clumpy far-downstream [\ion{S}{2}] emission structure. Such emission 
features set this cloud's emission properties apart from the those seen 
before in the well-studied, dense cloud regions of the Cygnus Loop. 

\subsubsection{X-ray Morphology and Radiative Shock Models} 

The X-ray morphology generally matches that seen in H$\alpha$. 
That is, one sees the X-ray emission 
trace the Balmer emission north to south (see
Fig.~\ref{fig:schmidtpspc}). 
In addition, the bright H$\alpha$ filaments 
south of the cloud correlate well with the extended X-ray emission 
located several arcminutes south of the cloud (Regions 5 and 6 of 
Fig.~\ref{fig:pspc}). 
This region itself may be another shock-cloud interaction, 
due to the clumpy appearance of the optical emission here. 
Regions 3 and 4 correspond to the north and south edges of the cloud,
where 
there is a brightening in the observed X-ray emission.

The fact that the X-ray emission arises along the edges of the cloud and
not from the interior of the cloud is puzzling. One possibility is that
there is a density gradient within the cloud. The lower-density component
of the cloud edges would not slow the shock below X-ray emitting temperatures,
while the higher-density core of the cloud would slow the shock. The existence
of X-rays along the edges of the cloud is in sharp contrast to other 
regions of the SNR, where the bulk of the X-ray emission arises from 
shocks reflected off of ISM clouds.

The X-ray spectral fits for Regions 3 and 4 give electron densities 
of $1.1$ and $0.7$ cm$^{-3}$ (assuming a line-of-sight depth of 
0.6 pc) and temperatures of 1.2 $\times$ 10$^6$ K. 
For complete electron-ion equilibration at this temperature, the
shock 
velocities are $\sim$ 290 km s$^{-1}$. Therefore, for the X-rays, the
ram 
pressure is $\sim$ 1.2 $\times$ 10$^{-9}$ dyne cm$^{-2}$, which
is consistent with both the X-ray emitting gas found in the XA region 
($\sim$ 9 $\times$ 10$^{-9}$ dyne cm$^{-2}$; \citealt{levenson01}) and the
southeast cloud ($\sim$ 10$^{-9}$ dyne cm$^{-2}$; \citealt{graham95}). It
is worth noting that the shock velocity derived from X-ray measurements
is $\sim$ 100 km s$^{-1}$ 
lower than what is found elsewhere in the remnant. However, the 
blowout region of the remnant may be evolving differently than the 
main SNR shell, so it is not surprising that the shock velocity here is 
different than elsewhere.

Our plots show several examples where the 
[\ion{O}{3}] seems to follow the H$\alpha$ by about 1\arcsec~ 
(e.g., Fig.~\ref{fig:regr3}). If the 
separation corresponds to the distance between the shock (going into 
partially neutral gas) and the peak of the [\ion{O}{3}] emissivity, 
then the shock velocity v$_s$ must be more than 120 km s$^{-1}$.
Models 
were run with a preshock neutral fraction of 0.5. The preshock 
neutral fraction cannot be larger than about 0.7 because the shock 
produced enough ionizing photons even when it is nonradiative to
ionize 
30\% of the H. On the other hand, 
the ionized fraction of H cannot be less than 0.3
without making the initial 
H$\alpha$ too weak compared with the [\ion{O}{3}] peak. A shock
speed of 
120 km s$^{-1}$ is too slow for this neutral fraction as might be 
expected from the lower effective shock speed with partially neutral 
preshock gas \citep{cox85}. An effective shock speed of 106 km
s$^{-1}$ will 
not completely ionize the gas beyond [\ion{O}{3}] so there is no
separate 
peak in emissivity behind the shock. 

We also computed the H$\alpha$ brightness assuming that 
the slit crosses a 1.5\arcsec~ high section of shock with a depth 
along the line-of-sight {\it l} = 5\arcmin~ = 0.63 pc (the size of our 
extraction regions in the X-ray analysis, and the extent of the filaments
in the plane of the sky).
We list the results from these models in Table~\ref{tab:radfits} for a 
distance of 440 pc, with the H$\alpha$--[\ion{O}{3}] separation designated as 
$\Delta$, the pre-shock number density expected from equal ram 
pressure with the X-ray gas as n$_{\rm cloud}$, and the H$\alpha$
brightness, assuming a preshock neutral fraction of 0.5, as 
I$_{\rm H\alpha}$.

From the values of $\Delta$ in Table~\ref{tab:radfits}, 
{\it V$_{shock}$} must be 
less than about 160 km s$^{-1}$ and the post-shock number density, 
n$_{\rm PS}$, must be about 3, implying that the ram pressure for the
cloud shock is slightly higher than that for the X-ray shock.
Considering that the X-ray gas 
is off to the sides of the optical cloud, this seems plausible. 
From I$_{{\rm H}\alpha}$, one concludes that the 
line of sight distance $l$ to produce the $\sim$ 1.5 $\times$
10$^{-15}$ 
erg cm$^{-2}$ s$^{-1}$ with n$_{\rm PS}$ 
$\sim$ 3 
must be smaller than the X-ray depth, not surprising 
given that this is a small filament, {\it l$_{{\rm H}\alpha}$} $<$ {\it
l$_{X}$} 
= 4 $\times$ 10$^{17}$ cm. 

Overall, our results are self-consistent within the constraints of
connecting 
optical and X-ray observations and the limits of the models. One 
expects that shocks faster than 150 km s$^{-1}$ would be thermally 
unstable \citep{innes92} 
so the higher velocity models listed in Table~\ref{tab:radfits} 
are not entirely 
reliable. But to first order, the cooling length of the 
gas should be approximately correct with the H$\alpha$ 
brightness given by the product of preshock 
neutral fraction and shock speed. The line of sight depth assumed 
for the X-ray emission also seems reasonable within the 
limitations of time-dependent ionization and depletion of 
refractory elements on grains. Both these effects these tend 
to decrease the emissivity, implying a density a little 
higher (no more than a factor of 2) than the numbers we derive. 

\section{CONCLUSIONS} 
\label{sect:conc} 

{\it ROSAT} X-ray data and ground based optical data for the 
southwestern region of the Cygnus Loop SNR show the early stages of
the 
interaction of a blast wave with a cloud in unprecedented detail. 
From our study of this shocked cloud, we conclude the following: 

1) The cloud began interacting with the shock $\sim 1200$ yr ago. 
This is supported 
by the lack of a standing bow shock behind the cloud, the lack of a
shock 
reconvergence point west of the cloud, and no evidence for 
instability formation along the edges of the cloud. 

2) The optical morphology of the cloud is substantially 
different than what is seen in 
the brighter regions of the remnant. Whereas many of the brighter
regions 
of the Cygnus Loop are the result of $\sim$ 400 km s$^{-1}$ shocks
hitting 
relatively higher density material, the low density and low shock
velocity 
nature of this region stretches the postshock cooling zone resulting in 
the diffuse [\ion{O}{3}] and clumpy [\ion{S}{2}] emissions
observed. 

3) The cloud's X-ray emission structure is also 
unlike that seen in the brighter 
optical and X-ray regions of the remnant. Little or no X-ray emission
is 
associated with the cloud itself, but there is bright X-ray emission 
associated with the northern and southern peripheries of the cloud. 
Furthermore, we 
derive a shock velocity of 290 km s$^{-1}$ which is significantly
lower than 
shock velocities found in other parts of the Cygnus Loop remnant. 

4) Small scale density fluctuations were found to exist within this ISM cloud 
which significantly altered the progression of the 
shock through the cloud. This is seen by the presence of
multiple small 
scale shocks which are seen throughout the cloud.

In summary, this relatively isolated, low density cloud in the
southwest 
limb of the Cygnus Loop has provided a 
revealing snapshot of the very early stages 
of a shock-cloud interaction. It shows how 
a cloud's initial density structure can strongly influence the observed
optical 
and kinematic morphology of the postshock gas. Further analyses of 
some of the exquisite details of 
this shock-cloud interaction may provide additional diagnostics with 
regards to two- and three-dimensional models of shocks overrunning
ISM 
clouds. 

\acknowledgements 

The authors wish to thank Ron Downes for assistance in the ground based 
observations, Bill Blair for helpful comments, and an 
anonymous referee for a careful reading of the manuscript. 
We also wish to thank the staff at the MDM 
Observatory for their consistently helpful and efficient assistance.
This work was partially supported by NASA through grant NAG 5-2088.

\onecolumn 

\begin{deluxetable}{lccccccc} 
\tablecaption{Measured Line Fluxes (H$\beta$ = 100)} 
\tablewidth{0pc} 
\tablehead{ 
\colhead{Line} & \colhead{$\lambda$} & \colhead{} & 
\colhead{} & \colhead{} & \colhead{F$_\lambda$} & \colhead{} &
\colhead{} \\ 
\cline{3-8} 
\colhead{} & \colhead{(\AA)} & \colhead{R1} & \colhead{R2} &
\colhead{S1\tablenotemark{a}} & \colhead{S2\tablenotemark{a}} &
\colhead{B1\tablenotemark{a}} & \colhead{B2} 
} 
\startdata 
H$\beta$ & 4861 & 100 & 100 & \nodata & \nodata & \nodata & 100 \\

$[{\rm O\, III}]$ & 4959 & 56 & 27 & 470 & 260 & \nodata & \nodata
\\ 
$[{\rm O\, III}]$ & 5007 & 150 & 98 & 1350 & 780 & 93 & 80 \\ 
$[{\rm O\, I}]$ & 6301 & 71 & 62 & \nodata & \nodata & \nodata &
\nodata \\ 
$[{\rm O\, I}]$ & 6365 & 19 & 22 & \nodata & \nodata & \nodata &
\nodata \\ 
$[{\rm N\, II}]$ & 6549 & 35 & 32 & \nodata & \nodata & \nodata &
\nodata \\ 
H$\alpha$ & 6563 & 338 & 317 & 300 & 300 & 300 & 340 \\ 
$[{\rm N\, II}]$ & 6584 & 104 & 100 & \nodata & \nodata & \nodata
& \nodata \\ 
$[{\rm S\, II}]$ & 6717 & 104 & 110 & 495 & 320 & \nodata &
\nodata \\ 
$[{\rm S\, II}]$ & 6731 & 75 & 78 & 360 & 240 & \nodata & \nodata
\\ 
F(H$\beta$)\tablenotemark{b} & & 8.0 & 7.9 & \nodata & \nodata &
\nodata & 0.4 \\ 
F(H$\alpha$)\tablenotemark{b} & & \nodata & \nodata & 0.9 & 1.5
& 2.6 & \nodata \\ 
\enddata 
\tablenotetext{a}{Fluxes are relative to H$\alpha$ = 300.} 
\tablenotetext{b}{In units of 10$^{-15}$ erg cm$^{-2}$ s$^{-1}$.}

\label{tab:opticalregions} 
\end{deluxetable} 

\begin{deluxetable}{cccccccc} 
\tablecolumns{8} 
\tablecaption{Model to the X-ray spectra of the southwest cloud and vicinity} 
\tablewidth{0pc} 
\tablehead{ 
\colhead{Region} & \colhead{$\alpha$ (J2000)} & \colhead{$\delta$
(J2000)} 
& \colhead{kT} & \colhead{N$_H$} & \colhead{F$_{\rm x}$\tablenotemark{a}} & 
\colhead{EM} & \colhead{$\chi^2$ (d.o.f)\tablenotemark{b}} \\ 
\colhead{} & \colhead{~h ~m ~s.s} & \colhead{\degr~ ~\arcmin~
~\arcsec} & 
\colhead{(keV)} & \colhead{10$^{20}$ atoms cm$^{-3}$} & 
\colhead{(0.1--1.0 keV)} & \colhead{cm$^{-6}$ pc} & \colhead{}} 
\startdata  
1 & 20 48 13.2 & 29 29 42 & 0.13$^{+0.01}_{-0.04}$ & 1.6$^{+2.8}_{-0.7}$ & 0.5 & 0.45 & 12.5 (11) \\ 
2 & 20 48 13.2 & 29 24 42 & 0.12$^{+0.06}_{-0.03}$ & 3.1$^{+5.5}_{-1.6}$ & 0.5 & 0.32 & 8.91 (6) \\ 
3 & 20 48 04.0 & 29 19 37 & 0.11$^{+0.03}_{-0.03}$ & 2.8$^{+0.4}_{-0.4}$ & 0.7 & 0.60 & 10.3 (6) \\ 
4 & 20 47 52.1 & 29 13 07 & 0.11$^{+0.05}_{-0.04}$ & 1.6$^{+0.4}_{-0.3}$ & 0.4 & 0.45 & 7.06 (5) \\ 
5 & 20 47 46.4 & 29 08 07 & 0.13$^{+0.05}_{-0.04}$ & 1.6$^{+0.4}_{-0.5}$ & 1.0 & 0.99 & 6.02 (5) \\ 
6 & 20 47 46.4 & 29 02 45 & 0.14$^{+0.02}_{-0.04}$ & 1.6$^{+0.8}_{-0.6}$ & 1.0 & 0.74 & 20.2 (11) \\ 
\enddata 
\tablenotetext{a}{In units of 10$^{-12}$ erg cm$^{-2}$ s$^{-1}$.}
\tablenotetext{b}{Statistical errors at 90\% confidence.}
\label{tab:regions} 
\end{deluxetable}

\begin{deluxetable}{ccccc} 
\tablecolumns{5} 
\tablecaption{Model fits to the radiative shocks seen in the southwest
Cloud} 
\tablewidth{0pc} 
\tablehead{ 
\colhead{V$_{\rm shock}$} & \colhead{$\Delta$} &\colhead{n$_{\rm cloud}$} & 
\colhead{I$_{\rm H\alpha}$} & \colhead{P$_{\rm opt}$ / P$_{\rm X-ray}$ 
\tablenotemark{a}} \\ 
\colhead{km s$^{-1}$ } & \colhead{\arcsec / n$_{\rm PS}$} & \colhead{cm$^{-3}$} & \colhead{10$^{-15}$ erg s$^{-1}$ cm$^{-2}$ $\times$ n$_{\rm PS}$} & \colhead{}}
\startdata 
140 & 4 & 1.0 & 3.5 & 0.75 \\ 
150 & 5 & 0.87 & 3.8 & 0.86 \\ 
160 & 8 & 0.76 & 4.0 & 0.97 \\ 
170 & 11 & 0.68 & 4.2 & 1.1 \\ 
180 & 16 & 0.60 & 4.5 & 1.2 \\ 
190 & 26 & 0.54 & 4.8 & 1.4 \\ 
200 & 40 & 0.50 & 5.0 & 1.5 \\ 
\enddata 
\tablenotetext{a}{Assuming n$_{\rm PS}$ = 3 cm$^{-3}$, 
and P$_{\rm X-ray}$ = 1.2 $\times$ 10$^{-9}$ dyne cm$^{-2}$.}
\label{tab:radfits} 
\end{deluxetable} 

\clearpage 

\clearpage 

\end{document}